\documentclass[reprint,prl,twocolumn,superscriptaddress,showpacs,showkeys,floatfix,preprintnumbers]{revtex4-1}
\usepackage{braket}
\usepackage{xargs}
\usepackage{mathtools}
\usepackage[english]{babel}
\usepackage[utf8]{inputenc}
\usepackage{amsmath,amssymb,braket,slashed,mathrsfs}
\usepackage{pifont}
\usepackage{MnSymbol}
\usepackage{subcaption}
\usepackage{graphicx}
\usepackage{tikz}
\usetikzlibrary{shapes,arrows,positioning}
\tikzset{decision/.style={diamond, draw, fill=blue!20, text width=4.5em, text badly centered, inner sep=0pt}}
\tikzset{block/.style={rectangle, draw, fill=blue!20, text width=10em, text centered, rounded corners, minimum width=3.5cm}}
\tikzset{block1/.style={rectangle, draw, fill=blue!20, text width=18.5em, text centered, rounded corners, minimum width=3.5cm}}
\tikzset{line/.style={draw, -latex, thick}}
\usepackage{color,hyperref}
\usepackage{multirow,array}
\usepackage{physics}

\usepackage{cleveref}
\newcommand{\be}{\begin{equation}}
	\newcommand{\ee}{\end{equation}}
\newcommand{\ba}{\begin{eqnarray}}
\newcommand{\ea}{\end{eqnarray}}

\newcommand{\nn}{\nonumber}

\newcommand{\innovation}{Collaborative Innovation Center of Quantum Matter, Beijing 100871, China}
\newcommand{\chep}{Center for High Energy Physics, Peking University, Beijing 100871, China}
\newcommand{\pkuphy}{School of Physics, Peking University, Beijing 100871,
	China}

\newcommand{\Uconn}{Department of Physics, University of Connecticut, Storrs, CT 06269, USA}

\newcommand{\Bonn}{Universit\"at Bonn, Helmholtz-Institut f\"ur Strahlen- und Kernphysik and Bethe Center for Theoretical Physics, D-53115 Bonn, Germany}
\newcommand{\FZJ}{Forschungszentrum J\"ulich, Institute for Advanced Simulation (IAS-4), D-52425 J\"ulich, Germany}

\newcommand{\Peng}{Peng Huanwu Collaborative Center for Research and Education, International Institute for Interdisciplinary and Frontiers, Beihang University, Beijing 100191, China}

\begin{document}

\captionsetup[figure]{justification=raggedright,singlelinecheck=false} % 让 figure 的 caption 左对齐
\captionsetup[table]{justification=raggedright,singlelinecheck=false}
\captionsetup[subfigure]{justification=centering}

	\title{Lattice QCD Study of Pion Electroproduction and Weak Production from a Nucleon}
	
	\author{Yu-Sheng~Gao}\affiliation{\pkuphy}
	\author{Zhao-Long~Zhang}\affiliation{\pkuphy}
        \author{Xu~Feng}\affiliation{\pkuphy}\affiliation{\innovation}\affiliation{\chep}
        \author{Lu-Chang Jin}\affiliation{\Uconn}
        \author{Chuan~Liu}\affiliation{\pkuphy}\affiliation{\innovation}\affiliation{\chep}
	\author{Ulf-G. Mei{\ss}ner}\affiliation{\Bonn}\affiliation{\FZJ}\affiliation{\Peng}
	
	\date{\today}
	
	\begin{abstract}
	Quantum fluctuations in QCD influence nucleon structure and interactions, with pion production serving as a key probe of chiral dynamics. 
	In this study, we present a lattice QCD calculation of multipole amplitudes at threshold, related to both pion electroproduction and weak production from a nucleon, 
	using two gauge ensembles near the physical pion mass. 
	We develop a technique for spin projection and construct multiple operators for analyzing the generalized eigenvalue problem 
	in both the nucleon-pion system in the center-of-mass frame and the nucleon system with nonzero momentum.
	The numerical lattice results are then compared with those extracted from experimental data and predicted by low-energy theorems incorporating 
	one-loop corrections.

	\end{abstract}
	
	\maketitle

	\section{Introduction}
	Quantum fluctuations are fundamental to modern physics, shaping many key phenomena.
	In QCD, gluon field  fluctuations are key to quark confinement and asymptotic freedom. 
	In the non-perturbative regime, they influence quark and gluon dynamics within nucleons, affecting their distributions 
	and interactions. These effects, in turn, shape how nucleons respond to external probes like photons and weak bosons. 
	When energy allows, quantum fluctuations can manifest as real particles, such as pions, in electroproduction and weak production. 
	Pion production is of particular interest, as pions, being Nambu-Goldstone boson of QCD, 
	reflect spontaneous chiral symmetry breaking and play a crucial role in chiral dynamics (see, e.g.,~\cite{Bernstein:2009dc}).

        The study of pion production through electromagnetic interactions has a long history. Low-energy theorems (LETs) successfully described charged pion photoproduction but initially failed for the
        $\gamma p\to\pi^0p$ process~\cite{Walker:1963zzb,Rossi:1973wf,Salomon:1983xn,Mazzucato:1986dz,Beck:1990da,Drechsel:1992pn}.
        Bernard et al. resolved these discrepancies by incorporating chiral perturbation theory (ChPT) corrections~\cite{Bernard:1991rt,Bernard:1992nc,Bernard:1994gm,Bernard:1995cj,Bernard:1996ti,Bernard:2001gz},
        advancing our understanding of QCD chiral dynamics, where quantum fluctuations (i.e, pion loops) are of prime importance.
        Increasing beam energies in recent electron-nucleon experiments have made it more challenging to probe pion production directly in the
        threshold region. The latest photoproduction data from MAMI, over a decade old~\cite{A2:2012lnr}, only cover energies above
        the second threshold, where the $\pi^+n$ channel opens. For a recent review, see~\cite{Meissner:2022yyi}.
        For electroproduction, where the photon carries nonzero four-momentum squared, the situation is less clear. 
        While the extension of the LETs to electroproduction and the ChPT analyses of experimental data exist~\cite{Bernard:1992ms,Bernard:1992rf,Bernard:1993bq,Bernard:1994dt,Bernard:1996bi},
        discrepancies among different measurements and deviations from ChPT predictions persist~\cite{Lindgren:2013eta}, 
        prompting further theoretical efforts~\cite{Hilt:2013fda,Fernandez-Ramirez:2012vlt,Rijneveen:2021bfw}.
        Lattice QCD calculations offer a first-principles approach to predicting threshold pion electroproduction, enabling direct comparisons with experiment and ChPT.
        Such comparisons are essential for improving our understanding of chiral dynamics in QCD.

	Weak pion production is crucial for neutrino oscillation experiments,
	where neutrino-nucleus interactions are a major source of systematic uncertainty.
	This impacts both intermediate-energy experiments like LBNF/DUNE~\cite{DUNE:2015lol}, 
	HyperK~\cite{Hyper-Kamiokande:2018ofw}, and JUNO~\cite{JUNO:2015zny}, as well as low-energy 
	coherent neutrino scattering programs~\cite{COHERENT:2017ipa,COHERENT:2020iec}. 
	The DUNE Conceptual Design Report highlights that uncertainties exceeding 1\% for signals and 5\% for backgrounds 
	could significantly reduce sensitivity to CP violation and the neutrino mass hierarchy~\cite{DUNE:2015lol}.
	A significant portion of the DUNE neutrino flux lies above the pion production threshold, making precise theoretical understanding of 
	pion production processes crucial. To achieve few-percent overall cross-section uncertainties, these processes must be understood at the 
	ten-percent level~\cite{Ruso:2022qes}. New data on neutrino scattering off proton or deuteron targets would provide valuable constraints.
	Theoretically, LETs are detailed in~\cite{Bernard:1993xh}. 
	While lattice QCD can offer crucial insights, current studies of matrix elements involving nucleon-pion rescattering states remain 
	exploratory compared to axial form factor calculations. Theoretical and computational advances are required to deliver results 
	with fully quantified uncertainties.	

	Lattice QCD calculations involving baryonic multi-hadron states are inherently challenging due to increased system complexity, 
	poorer signal-to-noise ratios, and potentially significant excited-state 
	contamination. Recent studies have explored the excited-state contamination in nucleon matrix
	elements~\cite{Barca:2022uhi,Grebe:2023tfx,Alexandrou:2024tin,Barca:2024hrl,Hackl:2024whw,Sasaki:2025qro}.  
	Building on our previous studies of nucleon electric polarizabilities~\cite{Wang:2023omf} and subtraction functions in
	forward Compton scattering~\cite{Fu:2024gxq}, we present a lattice QCD calculation of $\gamma^* N\to N\pi$, $W^* N\to N\pi$ 
	and $Z^* N\to N\pi$ matrix elements at the pion production threshold. Using two gauge ensembles near the physical pion mass but with different lattice spacings,
	we extract the multipole amplitudes $E_{0+}$ and $L_{0+}$ from pion electroproduction 
	and $L_{0+}^{(W)}$, $M_{0+}$ and $H_{0+}$ from weak production. A detailed comparison is conducted between lattice results, experimental data and 
	LET predictions.
	
	\section{Multipole amplitudes at the pion production threshold}
	Consider the process of $\gamma^*(k)+ N(p_1)\to N(p_2)+\pi(q)$, where $N$ represents a nucleon 
	(proton or neutron), $\pi$ denotes a pion, and $\gamma^*$ is a virtual photon with spacelike momenta if $k^2<0$. 
	Replacing $\gamma^*$ with $W^*$ or $Z^*$ transforms the electromagnetic process to a weak one.
	The transition matrix elements for the electromagnetic and axial weak current are given by
	\ba
	&&\mathcal{J}_\mu^{em}=\langle N\pi|J_\mu^{em}(0)|N\rangle,
	\nn\\
	&&\mathcal{J}_\mu^{W(Z),A}=\langle N\pi|J_\mu^{W(Z),A}(0)|N\rangle,
	\ea
	where the currents are defined as 
	\ba
	\label{eq:current}
	&&J_\mu^{em}=e\left(\frac{2}{3}\bar{u}\gamma_\mu u-\frac{1}{3}\bar{d}\gamma_\mu d\right), 
	\nn\\
	&&J_\mu^{W,A}=-\frac{g_2}{2\sqrt{2}}\bar{d}\gamma_\mu\gamma_5 u,
	\nn\\
	&&J_\mu^{Z,A}=-\frac{g_2}{4\cos\theta_W}\left(\bar{u}\gamma_\mu\gamma_5 u-\bar{d}\gamma_\mu \gamma_5d\right).
	\ea
	Here, $u,d$ are the up and down quark fields, $e$ and $g_2$ are the electromagentic and 
	weak $SU(2)_L$ coupling constants, and $\theta_W$ is the Weinberg angle. 
	The minus sign associated with the axial vector current reflects the $V-A$ structure of the weak interaction.
	
	In the $N\pi$ center-of-mass frame at threshold ($\vec{q}=\vec{p}_2=\vec{0}$), the electromagnetic current matrix element can be 
	expressed in terms of two S-wave multipole amplitudes, $E_{0+}$ and $L_{0+}$~\cite{Bernard:1993bq}
	\be
	\label{eq:extraction_L_E}
	[\vec{\mathcal{J}}^{em}]_{s',s}=\alpha_m
	\xi_{s'}^\dagger\left\{L_{0+}\,\hat{k}(\vec{\sigma}\cdot\hat{k})+E_{0+}\left[\vec{\sigma}-\hat{k}(\vec{\sigma}\cdot\hat{k})\right]\right\}\xi_s,
	\ee
	where $\alpha_m=8\pi\,i\left(m+M_\pi\right)$, with $m$ and $M_\pi$ being the masses of the nucleon and pion, respectively.
	$\xi_{s',s}$ are two-component Pauli spinors for the nucleon, normalized as $\xi_{s'}^\dagger \xi_s=\delta_{s,s'}$, 
	where $s$ and $s'$ denote the nucleon spin in the initial and final states.
	The multipole amplitude $E_{0+}$ characterizes the transverse coupling of the virtual photon to the nucleon spin, 
	while $L_{0+}$ characterizes the longitudinal coupling.
	Eq.~(\ref{eq:extraction_L_E}) applies for $\vec{k}\neq\vec{0}$. At $\vec{k}=\vec{0}$, $L_{0+}$ and $E_{0+}$ are equal in magnitude~\cite{Bernard:1993bq}, simplifying the expression to
	$[\vec{\mathcal{J}}^{em}]_{s',s}=\alpha_m E_{0+}\xi_{s'}^\dagger\vec{\sigma}\xi_s$.
	When $\vec{k}\neq\vec{0}$, $L_{0+}$ can also be extracted from the time component of the current, $\mathcal{J}_0^{em}$, using the Ward identity 
	\be
	[\mathcal{J}_0^{em}]_{s',s}=\alpha_m\frac{|\vec{k}|}{k_0}\xi_{s'}^\dagger \left(\vec{\sigma}\cdot\hat{k}\right)\xi_s\,L_{0+}.
	\ee
	
	For weak interactions mediated by the $W$ boson, the axial weak current matrix element is expressed as~\cite{Bernard:1993xh}
	\ba
	&&[\mathcal{J}_0^{W,A}]_{s',s}=\alpha_m\xi_{s'}^\dagger \xi_s\left(L_{0+}^{(W)}+\frac{k_0}{m}H_{0+}\right),
	\nn\\
	&&[\vec{\mathcal{J}}^{W,A}]_{s',s}=\alpha_m\xi_{s'}^\dagger \left(\frac{\vec{k}}{m}H_{0+}-i\left(\vec{\sigma}\times\hat{k}\right)M_{0+}\right)\xi_s,
	\ea
	where $L_{0+}^{(W)}$, $H_{0+}$ and $M_{0+}$ are the S-wave multipole amplitudes. To distinguish between electromagnetic and weak transitions, 
	the superscript $(W)$ is added to $L_{0+}$ for the weak transition. 
	For the $Z$ boson, the matrix elements are defined analogously, with the superscript $W$ replaced by $Z$.	
	These multipole amplitudes can also be expressed in the isospin basis. 
	The relationships between the multipole amplitudes in the physical and isospin bases are provided in the Supplemental Material~\cite{SM}.

	\section{Spin projection for correlation functions}
	
	The nucleon-pion operators $O_{N\pi}^{I,I_z}$ with isospin $I=\frac{1}{2}$ and $\frac{3}{2}$ are constructed using the isospin-triplet operator for the pion, $O_\pi^{I,I_z}$, 
	and the doublet operator for the nucleon, $O_N^{I,I_z}$,
	with appropriate coefficients. More details of the construction are given in the Supplemental Material~\cite{SM}. 
	
	To simplify the analysis, we first apply the projection operator $\mathcal{P}_+=\frac{1+\gamma_0}{2}$ to the nucleon field, reducing it to a two-component field. 
	Consequently, the spin structure of the correlation functions and matrix elements is analyzed within a $2\times2$ spin space. 
	The overlap of the interpolating operators $O_N^{I,I_z}$ and $O_{N\pi}^{I,I_z}$ with the nucleon and nucleon-pion ground state is expressed as
	\ba
	\label{eq:overlap}
	&&\langle 0|O_N^{I,I_z}(\vec{p},t)|N^{I,I_z},\vec{p},s\rangle_V=L^3\,Z_{N,L}(\vec{p})e^{-Et}\xi_s,
	\nn\\
	&&\langle 0| O_{N\pi}^{I,I_z}(\vec{p},t)|(N\pi)^{I,I_z},G_1^-,s\rangle_V=L^3\,Z_{N\pi,L}^I(\vec{p})e^{-E_{N\pi}^It}\xi_s,
	\nn\\
	\ea
	where the operators $O_N^{I,I_z}(\vec{p},t)$ and $ O_{N\pi}^{I,I_z}(\vec{p},t)$ are defined as
	\ba
	&&O_N^{I,I_z}(\vec{p},t)=\sum_{\vec{x}}O_N^{I,I_z}(\vec{x},t)e^{i\vec{p}\cdot\vec{x}},
	\nn\\
	&&O_{N\pi}^{I,I_z}(\vec{p},t)=\frac{1}{N_R}\sum_{\hat{R}\in O_h}\sum_{\vec{x},\vec{y}}O_{N\pi}^{I,I_z}(\vec{x},\vec{y},t)e^{i\hat{R}\vec{k}\cdot(\vec{x}-\vec{y})},
	\ea
	with $N_R=\sum_{\hat{R}\in O_h}1$, and $\hat{R}$ is an element of the hypercubic group $O_h$, which describes the rotational symmetry in a finite volume.
	The operator $ O_{N\pi}^{I,I_z}(\vec{p},t)$ is defined in the center-of-mass frame, with $\vec{p}$ serving only as an indicator of the operator construction.
	The coordinates $\vec{x}$ and $\vec{y}$ correspond to the spatial positions of the nucleon and pion operators, respectively.
	The factor $L^3$ accounts for the finite volume, where $L$ represents the lattice size.
	The subscript $V$ denotes the states in a finite volume. $E$ and $E_{N\pi}^I$ represent the energies of the nucleon and nucleon-pion states, respectively.

	The $G_1^-$ representation is a two-dimensional irreducible representation of $O_h$, 
	with basis states labeled by the index $s$, which remain consistent with the spin index in the infinite-volume limit.
	The finite-volume states are normalized by
	\ba
	&&{_V}\langle N^{I,I_z'},\vec{p}',s'|N^{I,I_z},\vec{p},s\rangle_V=2EL^3\delta_{I_z,I_z'}\delta_{s,s'}\delta_{\vec{p},\vec{p}'},
	\nn\\
	&&{_V}\langle (N\pi)^{I,I_z'},G_1^-,s'|(N\pi)^{I,I_z},G_1^-,s\rangle_V=2E_{N\pi}^I L^3\delta_{I_z,I_z'}\delta_{s,s'}.
	\nn\\
	\ea

	The correlation functions are constructed as
	\ba
	C_{N\pi JN}(t_f,t_J,t_i)&=&\langle O_{N\pi}^{I'',I_z''} (\vec{0},t_f)\tilde{J}_\mu^{I',I_z'}(\vec{k},t_J) \bar{O}_N^{I,I_z}(\vec{p},t_i)\rangle,
	\nn\\
	C_{N\pi}(t_f,t_i)&=&\langle O_{N\pi}^{I,I_z} (\vec{0},t_f)\bar{O}_{N\pi}^{I,I_z}(\vec{0},t_i)\rangle,
	\nn\\
	C_{N}(t_f,t_i)&=&\langle O_{N}^{I,I_z} (-\vec{p},t_f)\bar{O}_{N}^{I,I_z}(\vec{p},t_i)\rangle.
	\ea
	According to Eq.~(\ref{eq:overlap}), a typical operator $O^{I,I_z}$ acts as an annihilation operator, removing a state with isospin $(I,I_z)$. 
	However, the same operator can also act as a creation operator, generating a state with isospin $(I,-I_z)$.
	To clarify this distinction, we introduce the notation $\tilde{O}^{I,-I_z}$ to represent $O^{I,I_z}$ when it functions as a creation operator. This convention similarly applies to the current operator.
	By using $\tilde{J}_\mu^{I',I_z'}$ in the correlation function, we ensure that the isospin relation $I_z+I_z'=I_z''$ holds. Specifically, we set $(I_z,I_z')=(\frac{1}{2},0)$. Other choices of $(I_z,I_z')$ can be
	related to this setup through the Wigner-Eckart theorem. Additionally, the current’s momentum
	$\vec{k}$ satisfies the momentum conservation condition $\vec{p} + \vec{k} = \vec{0}$.

	For the correlation functions $C_{N}(t_f,t_i)$ and $C_{N\pi}(t_f,t_i)$, at large time separation $t_f-t_i$, we obtain
	\ba
	\label{eq:t_dep_1}
	\frac{1}{2}\operatorname{Tr}[C_{N}(t_f,t_i)]&=&L^3\frac{Z_{N,L}(\vec{p})^2}{2E}e^{-E(t_f-t_i)},
	\nn\\
	\frac{1}{2}\operatorname{Tr}[C_{N\pi}(t_f,t_i)]&=&L^3\frac{Z_{N\pi,L}^I(\vec{0})^2}{2E_{N\pi}^I}e^{-E_{N\pi}^I(t_f-t_i)}.
	\ea
	For the correlation function $C_{N\pi JN}$, we use $\vec{C}$ and $C^0$ to distinguish its spatial and temporal components, 
	and denote vector and axial-vector current insertions by $J=V$ and $A$, respectively. Before applying the trace operator, we first express
	\be
	\label{eq:four_point_func}
	C_{N\pi JN}=\alpha_{N\pi JN}
	\sum_{s',s}\xi_{s'}[\mathcal{J}]_{s',s}\xi_s^\dagger,
	\ee
	where the coefficient is given by
	\be
	\alpha_{N\pi JN}=L^3\frac{Z_{N\pi,L}^I(\vec{0})}{2E_{N\pi}^I}e^{-E_{N\pi}(t_f-t_J)}
	\frac{Z_{N,L}(\vec{p})}{2E}e^{-E(t_J-t_i)}.
	\ee
	
	To extract the five multipole amplitudes, we define the following spin projection operators
	\ba
	&&\mathcal{P}_{L_{0+}}=\frac{k_0}{|\vec{k}|}(\hat{k}\cdot\vec{\sigma}),\quad \vec{\mathcal{P}}_{E_{0+}}=\frac{1}{2}(\vec{\sigma}-(\hat{k}\cdot\vec{\sigma})\hat{k}),
	\nn\\
	&&\vec{\mathcal{P}}_{H_{0+}}=\frac{m}{|\vec{k}|}\hat{k},\quad\mathcal{P}_{L_{0+}^{(W)},H_{0+}}=1,
	\quad \vec{\mathcal{P}}_{M_{0+}}=\frac{i}{2}(\vec{\sigma}\times \hat{k}).
	\ea
	These projection operators are applied to correlation functions to extract the multipole amplitudes. For example, 
	\ba
	\label{eq:t_dep_2}
	&&\frac{1}{N_R}\sum_{\hat{R}\in O_h}\frac{1}{2}\operatorname{Tr}\left[\mathcal{P}_{L_{0+}}\cdot C_{N\pi VN}^0\right]\Big|_{\hat{R}\vec{k}}
	\nn\\
	&&\hspace{1cm}=\left(2\mu_{N\pi}\right)^{-\frac{1}{2}}f_{LL}^{-{\frac{1}{2}}}\alpha_{N\pi JN}\alpha_m L_{0+},
	\ea
	where $f_{LL}$ is the Lellouch-L\"uscher factor that relates the finite-volume state to the infinite-volume one, and $\mu_{N\pi}=\frac{m M_\pi}{E_{N\pi}}$ is the reduced mass of the nucleon-pion system.
	Other multipole amplitudes can be extracted by applying the corresponding spin projection operators to the relevant correlation functions, as summarized in Table~\ref{tab:multiple_amp}.
	These projection operators are valid for $\vec{k}\neq\vec{0}$. For $\vec{k}=\vec{0}$, $E_{0+}$ can be extracted by applying $\vec{\mathcal{P}}_{E_{0+}}=\vec{\sigma}/3$ to $\vec{C}_{N\pi VN}$.

	\begin{table}[htbp]
	\small
	\centering
	\begin{tabular}{ccccc}
	\hline
	\hline
	  $\mathcal{P}_{L_{0+}}$ & $\vec{\mathcal{P}}_{E_{0+}}$ & $\vec{\mathcal{P}}_{H_{0+}}$ & $\mathcal{P}_{L_{0+}^{(W)},H_{0+}}$ & $\vec{\mathcal{P}}_{M_{0+}}$ \\
 	\hline
	  $C_{N\pi VN}^0$ & $\vec{C}_{N\pi VN}$ & $\vec{C}_{N\pi AN}$ & $C^0_{N\pi AN}$ & $\vec{C}_{N\pi AN}$ \\
	\hline
	  $L_{0+}$ & $E_{0+}$ & $H_{0+}$ & $L_{0+}^{(W)}+\frac{k_0}{m}H_{0+}$ & $M_{0+}$ \\
	\hline
	\hline
	\end{tabular}%
	\caption{The correspondence between spin projection operators, correlation functions, and multipole amplitudes.}
	\label{tab:multiple_amp}%
	\end{table}%

	Applying the projection operator $\vec{\mathcal{P}}_{L_{0+}}=(\hat{k}\cdot\vec{\sigma})\hat{k}$ to $\vec{C}_{N\pi VN}$ can also extract $L_{0+}$,
	but $\mathcal{P}_{L_{0+}}$ is preferred as it reduces systematic effects. For instance, if the lattice size is tuned such that $k_0=0$, $L_{0+}$ should vanish,
	yet $\vec{\mathcal{P}}_{L_{0+}}$ fails to ensure this due to systematic uncertainties. In our study, the factor $k_0/|\vec{k}|$ takes small values of $0.376$, $0.170$ and $0.063$,
	which helps suppress systematic effects when using $\mathcal{P}_{L_{0+}}$. Conversely, we observe significant excited-state contamination when using $\vec{\mathcal{P}}_{L_{0+}}$,
	leading us to adopt $\mathcal{P}_{L_{0+}}$ exclusively.
	Further details on the design of spin operators and the discussion of Lellouch-L\"uscher factor are provided in the Supplemental Material~\cite{SM}.

	\section{Operator optimization}

	To reduce excited-state contamination, we use $O_{N\pi}^{I,I_z}(\vec{p},t)$ with the four lowest momentum modes
	\be
	\frac{\vec{p}}{2\pi/L}=(0,0,0),\, (0,0,1),\, (0,1,1),\, (1,1,1).
	\ee
	These operators are denoted as $O_{N\pi}^{(n)}$,
	where $n=1,2,3,4$ corresponds to increasing momentum modes. For simplicity, we have omitted the isospin index.
	Using these operators, we construct a $4\times4$ correlation function matrix with elements given by
	\be
	\label{eq:correlation_matrix_Npi}
	\mathcal{M}^{n,m}_{N\pi}(t_f-t_i)=\frac{1}{2}\operatorname{Tr}\langle O_{N\pi}^{(n)}(t_f) \bar{O}_{N\pi}^{(m)}(t_i)\rangle.
	\ee
	The four lowest states are defined as
	$|N\pi,G_1^-,s,n\rangle_V$ with $n=1,2,3,4$. 
	
	By solving the generalized eigenvalue problem (GEVP), we construct optimized $N\pi$ operators as
	\be
	\tilde{O}_{N\pi}=O_{N\pi}^{(1)}+c_2O_{N\pi}^{(2)}+c_3O_{N\pi}^{(3)}+c_4O_{N\pi}^{(4)},
	\ee
	where the coefficients $c_m$ are determined using the standard GEVP procedure~\cite{Luscher:1990ck,Blossier:2009kd}.
	Assuming that the lowest four states dominate the correlation function matrix, with higher excited-state contributions being negligible compared to the statistical noise,
	the coefficients $c_m$ can be considered to satisfy the condition
	\be
	\langle 0|\tilde{O}_{N\pi}|N\pi,G_1^-,s,n\rangle_V=0,\quad \mbox{for }n=2,3,4.
	\ee

	For the nucleon operator with nonzero momentum $\vec{p}$, parity is no longer a good quantum number, allowing mixing with
	$N\pi$ operators of the same momentum. We consider three operators
	\ba
	\label{eq:N_GEVP}
	&&O_N^{(1)}=O_N(\vec{p}),\quad O_N^{(2)}=(\hat{p}\cdot\vec{\sigma})O_N(\vec{p})O_\pi(\vec{0}),
	\nn\\
	&&O_N^{(3)}=(\hat{p}\cdot\vec{\sigma})O_N(\vec{0})O_\pi(\vec{p}),
	\ea
	using which, we construct the correlation function matrix
	\be
	\label{eq:correlation_matrix_N}
	\mathcal{M}^{n,m}_{N}(t_f-t_i)=\frac{1}{2}\operatorname{Tr}\langle O_{N}^{(n)}(t_f) \bar{O}_{N}^{(m)}(t_i)\rangle.
	\ee
	It is explained in the Supplemental Material why this matrix is suitable for GEVP analysis~\cite{SM}. 
	By solving the GEVP, we obtain the optimized nucleon operator
	\be
	\label{eq:nucleon_GEVP}
	\tilde{O}_{N}=O_{N}^{(1)}+d_2O_{N}^{(2)}+d_3O_{N}^{(3)}.
	\ee
	where the coefficients $d_{2,3}$ satisfy
	\be
	\langle 0|\tilde{O}_{N}|N(\vec{0})\pi(\vec{p})\rangle_V=0,\quad \langle 0|\tilde{O}_{N}|N(\vec{p})\pi(\vec{0})\rangle_V=0.
	\ee
	
	We construct the correlation function using the optimized operators
	\ba
	\tilde{C}_{N\pi JN}(t_f,t_J,t_i)=\langle \tilde{O}_{N\pi}(t_f)\tilde{J}_\mu(t_J)\bar{\tilde{O}}_N(t_i)\rangle,
	\ea
	where we include only terms from $C_{N\pi JN}$ and those proportional to the coefficients $c_{2,3,4}$ and $d_{2,3}$.
	Terms involving products of $c_m d_n$ are treated as higher-order corrections and neglected.
	A challenge in computing correlation functions like $\langle O_{N\pi}^{(1)}(t_f)\tilde{J}_\mu(t_J)\bar{O}_N^{(n)}(t_i)\rangle$
	for $n=2,3$ is the evaluation of five-point correlation functions. 
	Since the disconnected diagrams encapsulate the contributions from $\langle N(\vec{0})|N(\vec{0})\rangle \langle \pi({\vec{0}})|\tilde{J}_\mu|\pi(\vec{p})\rangle$
	and $\langle N(\vec{0})|\tilde{J}_\mu|N(\vec{p})\rangle \langle \pi({\vec{0}})|\pi(\vec{0})\rangle$, 
	which are enhanced by a factor of spatial volume of the lattice, they
	dominate the five-point correlation functions based on the factorization approximation 
	\ba
	&&\langle N(\vec{0})\pi({\vec{0}})|\tilde{J}_\mu|N(\vec{0})\pi(\vec{p})\rangle \approx \langle N(\vec{0})|N(\vec{0})\rangle \langle \pi({\vec{0}})|\tilde{J}_\mu|\pi(\vec{p})\rangle
	\nn\\
	&&\langle N(\vec{0})\pi({\vec{0}})|\tilde{J}_\mu|N(\vec{p})\pi(\vec{0})\rangle \approx \langle N(\vec{0})|\tilde{J}_\mu|N(\vec{p})\rangle \langle \pi({\vec{0}})|\pi(\vec{0})\rangle.
	\nn
	\ea
	Therefore, we compute only the disconnected contributions as approximations to the challenging five-point functions, which is feasible within current lattice QCD studies.
	Though certain simplifications have been made, they mainly affect corrections for excited-state contamination and are therefore acceptable. 
	Future work to develop methods for handling five-point correlation functions is beneficial.
	
	\section{Numerical analysis}

        We used two $2+1$-flavor domain wall fermion ensembles, 24D and 32Df, from the RBC-UKQCD Collaboration~\cite{RBC:2014ntl}, which have similar pion masses  
        (142.6(3) and 142.9(7) MeV~\cite{Lin:2024khg}), comparable spatial volumes ($L = 4.6$ fm), the same discretization but different lattice spacings ($a^{-1} = 1.023(2)$ and $1.378(5)$ GeV).
        For each configuration, we generate 1024 point-source and 1024 smeared-source propagators at randomly chosen spatiotemporal locations to compute the correlation functions, 
        using the random sparsening-field technique~\cite{Li:2020hbj,Detmold:2019fbk}. Smeared nucleon operators and local current operators are used, 
        with renormalization factors provided in Ref.~\cite{Feng:2021zek}.
        Additional details on the computation of four-point correlation functions can be 
        found in Refs.~\cite{Fu:2022fgh,Wang:2023omf,Ma:2023kfr}.
        
\begin{figure}[htb]
        \centering
        \includegraphics[width=0.48\textwidth,angle=0]{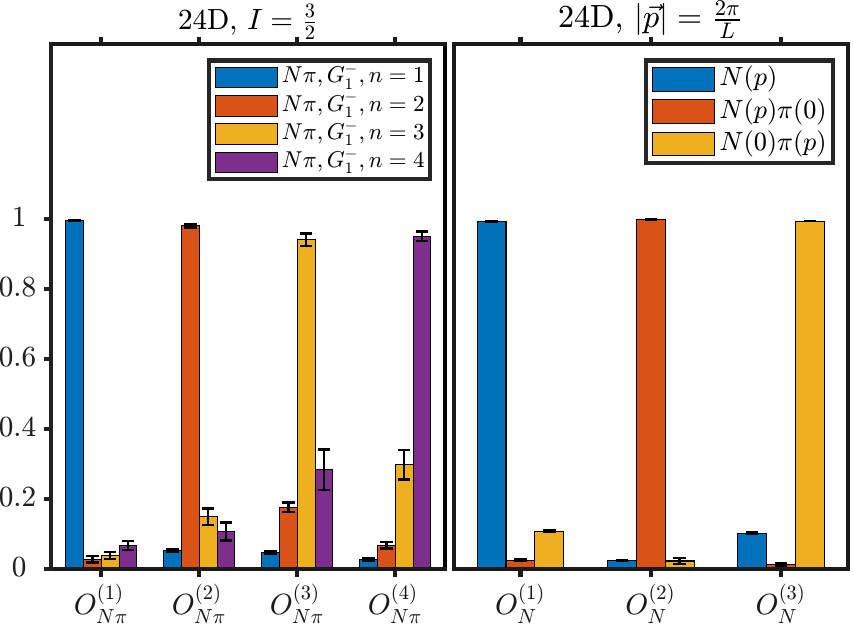}
        \caption{The overlap of the operators $O_{N\pi}^{(n)}$ for $n=1,2,3,4$ (left) and $O_N^{(n)}$ for $n=1,2,3$ (right) with the eigenstates from the GEVP analysis.}
        \label{fig:GEVP}
        \end{figure}

        Taking the 24D ensemble and the $I=3/2$ $N\pi$ system as an example, we show the GEVP analysis results in the left panel of Fig.~\ref{fig:GEVP}.
        For each operator $O_{N\pi}^{(n)}$, we plot its overlap with the state $|N\pi,G_1^-,s,m\rangle$, defined as $c_{n,m}=|\langle0|O_{N\pi}^{(n)}|N\pi,G_1^-,s,m\rangle_V|$, normalized 
        by $\sqrt{\sum_mc_{n,m}^2}$.
        Similarly, for the nucleon system with momentum $|\vec{p}|=\frac{2\pi}{L}$, the right panel of Fig.~\ref{fig:GEVP} presents the corresponding GEVP results. 
        The overlaps with excited states for the operators $O_{N\pi}^{(1)}$ and $O_N^{(1)}$ are about 5\% and 10\%, respectively.
        Although these overlaps are small, eliminating excited-state contamination remains crucial, as discussed below.
        
        Using Eqs.~(\ref{eq:t_dep_1}) and (\ref{eq:t_dep_2}), we extract the effective multipole amplitude for given time
        separations $t_{N\pi}-t_J$ and $t_J-t_N$. 
        Since the initial and final state operators differ significantly, we analyze their time dependence separately.
        We first fix $t_N-t_J$ and compute the multipole amplitude as a function of $t_{N\pi}-t_J$ to assess excited-state contamination on the $N\pi$ side. 
        By fitting this dependence, we obtain an effective multipole amplitude dependent only on $t_N-t_J$. We then examine excited-state contamination on the $N$ side 
        and extract the final multipole amplitude by fitting its $t_N-t_J$ dependence.

\begin{figure}[htb]
        \centering
        \includegraphics[width=0.48\textwidth,angle=0]{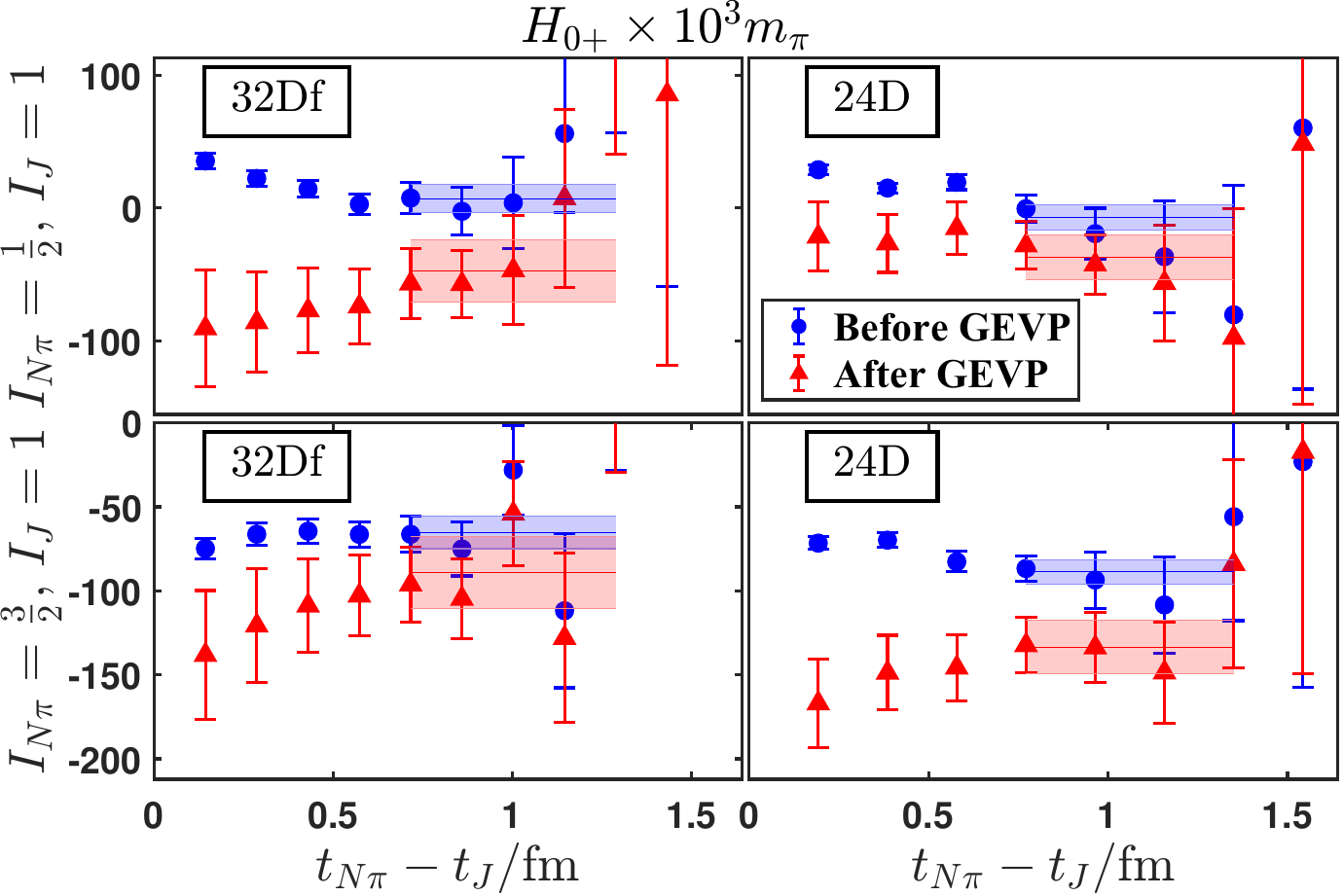}
        \caption{Effective multipole amplitude $H_{0+}$ as a function of $t_{N\pi}-t_J$ for $t_J-t_N=0.58$ fm. }
        \label{fig:H_Npi}
        \end{figure}

        We take $H_{0+}$ and $L_{0+}$ as examples to illustrate the impact of the GEVP correction on the $N\pi$ and $N$ sides, respectively. 
        Fig.~\ref{fig:H_Npi} shows $H_{0+}$ as a function of $t_{N\pi}-t_J$ for $t_J-t_N=0.58$ fm, comparing results with and without GEVP corrections.
        Here, ``with GEVP'' refers to the use of the optimized operator $\tilde{O}_{N\pi}(t)$, which retains both the $C_{N\pi JN}$ term and terms proportional to $c_{2,3,4}$,
        whereas ``without GEVP'' omits the $c_{2,3,4}$ corrections.
        As the coupling of $O_{N\pi}^{(1)}$ with excited states is weak, the GEVP corrections are not highly significant as shown in Fig~\ref{fig:H_Npi}.
        However, they still cause a shift by around 1-3 $\sigma$. 

\begin{figure}[htb]
        \centering
        \includegraphics[width=0.48\textwidth,angle=0]{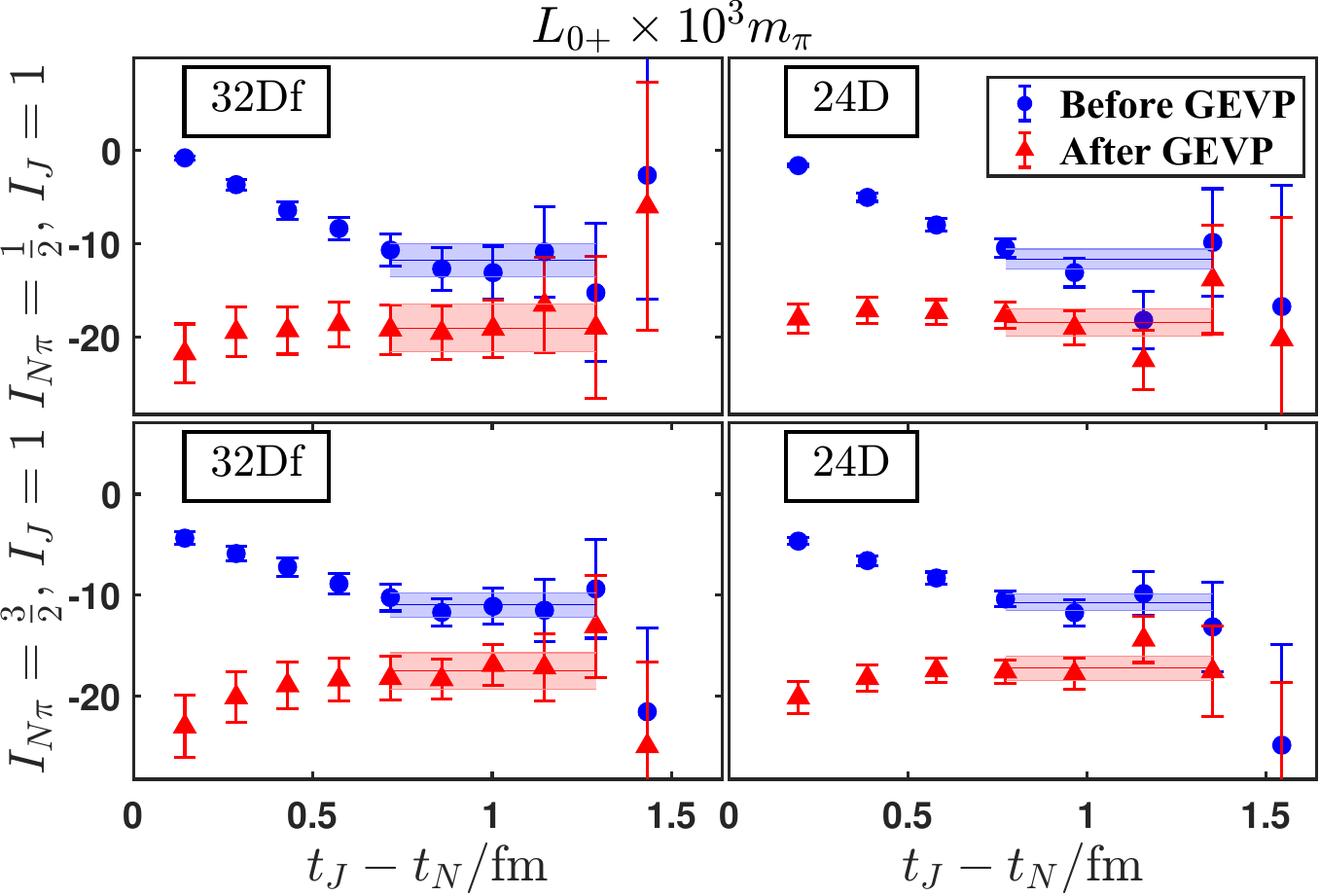}
         \caption{
Effective multipole amplitude $L_{0+}$ as a function of $t_J-t_N$, obtained from a fit using GEVP-corrected data on the $N\pi$ side to analyze the time dependence on the $N$ side, both before and after GEVP correction on the $N$ side.}
        \label{fig:L_N}
        \end{figure}

  	Fig.~\ref{fig:L_N} compares $L_{0+}$ as a function of $t_J-t_N$ with and without GEVP corrections on the $N$ side. 
        Here, the $t_J-t_N$ dependence 
        is analyzed using data that include the $c_{2,3,4}$ corrections.
        The terms ``with GEVP'' and ``without GEVP'' indicate whether the $d_{2,3}$ terms are included.
        The GEVP correction is crucial: before GEVP, significant excited-state contamination is visible at small $t_J-t_N$, whereas after correction, a clearer plateau emerges. 
        The difference can reach 6 $\sigma$ in the more precise 24D data. 
        Since nucleon operators with nonzero momenta are widely used in lattice calculations, such as for parton distribution functions, 
        removing excited-state contamination using techniques like GEVP is essential. 
	It is worth noting that GEVP optimization has only a mild effect on the single-nucleon two-point correlation function $C_N(t_f,t_i)$.
	However, for matrix elements extracted from $C_{N\pi JN}(t_f,t_J,t_i)$, the impact of GEVP is significant. A more detailed discussion is provided 
	in the Supplemental Material~\cite{SM}.
	Figures illustrating the fitting quality for other multipole amplitudes are also included in \cite{SM}.

        	\section{Results and conclusion}

        \begin{figure}[htb]
        \centering
        \includegraphics[width=0.48\textwidth,angle=0]{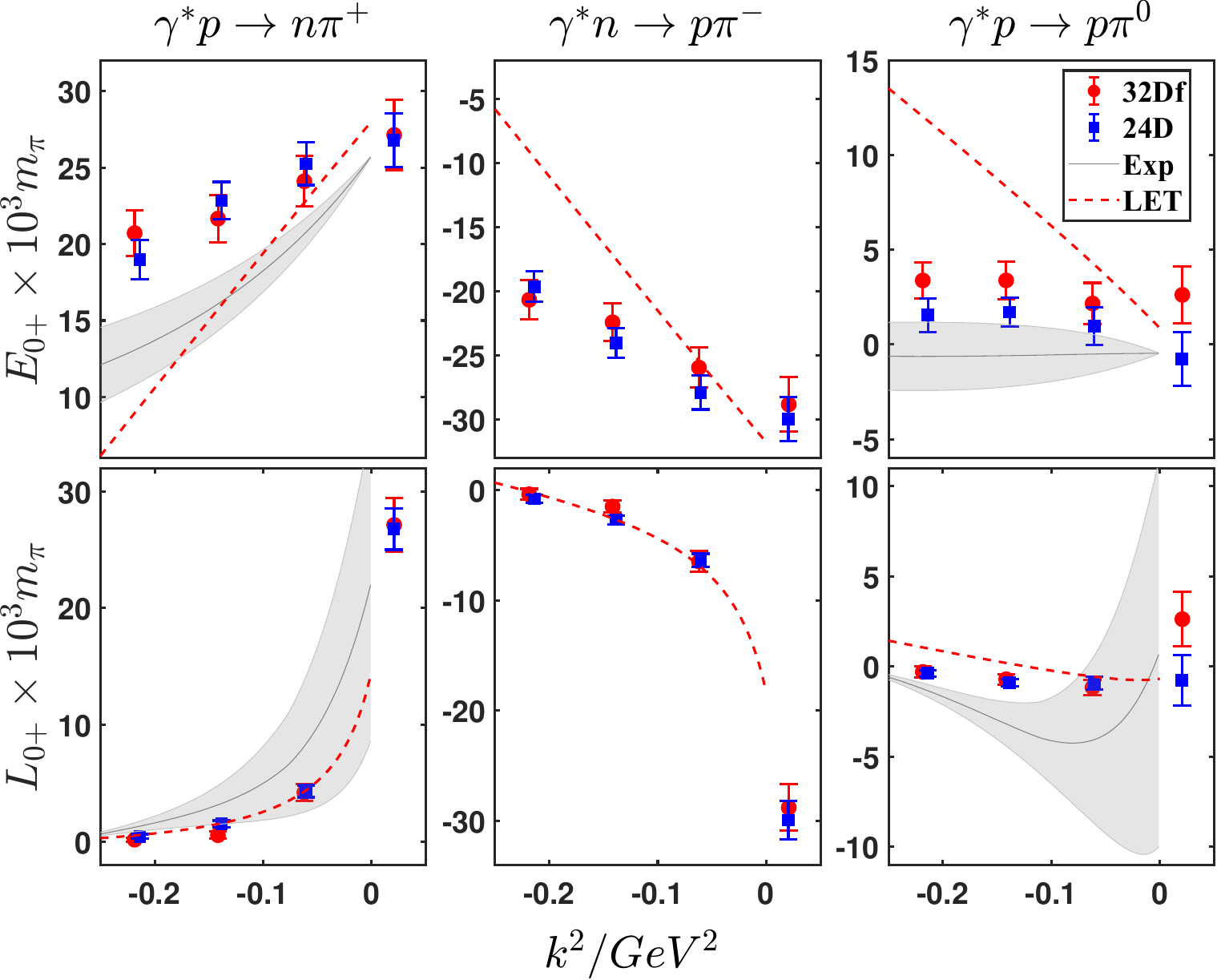}
        \caption{Comparison between lattice results for $L_{0+}$ and $E_{0+}$ with those extracted from experimental data and predicted by LETs. 
        The lattice value of $L_{0+}$ at $k^2=M_\pi^2$ is determined by enforcing the condition $L_{0+}=E_{0+}$ at $\vec{k}=\vec{0}$. 
        Experimental data and LET predictions are available in the spacelike region ($k^2<0$), as reported in Refs.~\cite{Mai:2023cbp} and \cite{Bernard:1993bq}.}
        \label{fig:comparison_phy_basis}
        \end{figure}	
        
        Fig.~\ref{fig:comparison_phy_basis} shows the momentum dependence of the multipole amplitudes obtained from fits to both the $t_{N\pi}-t_J$ and $t_J-t_N$ dependences.
        The amplitudes $L_{0+}$ and $E_{0+}$ are presented in the physical basis, allowing for a direct comparison of the lattice results with 
        extractions from experimental data and predictions from LETs, including $O((M_\pi/m)^2)$ corrections~\cite{Bernard:1993bq}.
        Lattice results for other multipole amplitudes in the isospin basis are provided in the Supplemental Material~\cite{SM}.
        Several partial-wave analyses based on experimental data exist~\cite{Kamano:2013iva,Nakamura:2015rta,Kamano:2016bgm,Briscoe:2023gmb,Mai:2021vsw,Mai:2021aui,Mai:2023cbp}. 
        In this work, we compare our results with the most recent analysis within a coupled-channel framework~\cite{Mai:2023cbp}.
        Note, however, this analysis does not incorporate matching to the ChPT amplitude at low energies and momenta,
        and only proton target data are analyzed, as the neutron in the initial state is bound in a deuteron or ${^3}$He, making the theoretical interpretation less clean.
        
        In Fig.~\ref{fig:comparison_phy_basis}, the lattice data exhibit a similar trend to both experimental analyses and LET predictions
        but align more closely with the experimental results while deviating more from LETs.
        This discrepancy arises because LETs omit higher-order corrections within the framework of ChPT.
	Although some differences exist between the lattice and experimental results, the large uncertainties in the experimental data imply that 
	the lattice and experimental results are either consistent or deviate by around 2-3 $\sigma$.
	Additionally, some deviations between the two lattice ensembles can be observed.
	While these differences may arise from lattice artifacts, they could also result from statistical fluctuations. 
	Further investigation into these effects, using larger statistics and additional lattice spacings, would be a valuable direction for future studies.
	Improved precision in both lattice calculations and experiments will enable 
        mutual validation and deepen our understanding of pion production from a nucleon, a fundamental quantum fluctuation process.
        
        As the first lattice QCD study of pion production, our approach to spin projection and multiple operator construction 
        extends techniques developed to study excited-state effects in nucleon observables 
        and holds promise for the eventual calculation of shallow inelastic neutrino-nucleon scattering.

\begin{acknowledgments}
{\bf Acknowledgments} --
        X.F. and L.C.J. gratefully acknowledge many helpful discussions with our colleagues from the RBC-UKQCD Collaborations. We would like to express our gratitude to T.-S. H. Lee and I. Strakovsky
        for the discussion on partial-wave analysis of the experimental data.
        X.F., Y.S.G., C.L. and Z.L.Z. were supported in part by NSFC of China under Grant No. 12125501, No. 12293060, No. 12293063 and No. 12141501.
        L.C.J. acknowledges support by DOE Office of Science Early Career Award No. DE-SC0021147 and DOE Award No. DE-SC0010339.
        The work of U.-G.M. was supported in part by the CAS President’s International Fellowship Initiative (PIFI) (Grant No. 2025PD0022) and by the Deutsche Forschungsgemeinschaft (DFG, German Research Foundation) as part of the CRC 1639 NuMeriQS – project no. 511713970.
	The research reported in this work was carried out using the computing facilities at Chinese National Supercomputer Center in Tianjin.
	It also made use of computing and long-term storage facilities of the USQCD Collaboration, which are funded by the Office of Science of the U.S. Department of Energy.
\end{acknowledgments}
	
	\bibliography{ref.bib}
	
\clearpage

\setcounter{page}{1}
\renewcommand{\thepage}{Supplemental Material -- S\arabic{page}}
\setcounter{table}{0}
\renewcommand{\thetable}{S\,\Roman{table}}
\setcounter{equation}{0}
\renewcommand{\theequation}{S\,\arabic{equation}}
\setcounter{figure}{0}
\renewcommand{\thefigure}{S\,\arabic{figure}}

\section{Supplemental Material}
	
	\subsection{Lattice operator conventions}
	
	In this calculation, we use Euclidean lattice interpolating operators for the pion and nucleon fields, defined as follows.
	For the pion fields, the operators are given by
	\ba
	&&O_{\pi^+}^{E}=\bar{d}\gamma_5^Eu,\quad \bar{O}_{\pi^+}^E=-\bar{u}\gamma_5^Ed,
	\nn\\
	&&O_{\pi^0}^E=\frac{1}{\sqrt{2}}(\bar{u}\gamma_5^Eu-\bar{d}\gamma_5^Ed),
	\quad \bar{O}_{\pi^0}^E=-\frac{1}{\sqrt{2}}(\bar{u}\gamma_5^Eu-\bar{d}\gamma_5^E d),
	\nn\\
	&&O_{\pi^-}^E=\bar{u}\gamma_5^E d,\quad \bar{O}_{\pi^-}^E=-\bar{d}\gamma_5^Eu,
	\ea
	For the nucleon fields, they are
	\ba
	O_p^E=\mathcal{P}_+\epsilon_{abc}u_a(u_b^TC\gamma_5^Ed_c),\quad O_n^E=\mathcal{P}_+\epsilon_{abc}d_a(u_b^TC\gamma_5^Ed_c),
	\nn\\
	\bar{O}_p^E=\epsilon_{abc}\bar{u}_a(\bar{u}_b^TC\gamma_5^E\bar{d}_c)\mathcal{P}_+,\quad \bar{O}_n^E=\epsilon_{abc}\bar{d}_a(\bar{u}_b^TC\gamma_5^E\bar{d}_c)\mathcal{P}_+.
	\nn\\
	\ea
	In the chiral representation, the charge conjugation matrix $C$ is given by $C=i\gamma_2^E\gamma_0^E$. 
	The gamma matrices in Euclidean space (denoted with a superscript $E$)
	are related to those in Minkowski space (without the superscript) by the following relations
	\be
	\gamma_0^E=\gamma^{0},\quad \gamma_i^E=-i\gamma^i,\quad \gamma_5^E=\gamma_0^E\gamma_1^E\gamma_2^E\gamma_3^E=\gamma_5.
	\ee 
	The Minkowski gamma matrices are defined as
	\be
	\gamma^0=\begin{pmatrix} 0 & 1 \\ 1 & 0 \end{pmatrix},\quad \gamma^i=\begin{pmatrix} 0 & \sigma^i \\ -\sigma^i & 0 \end{pmatrix},\quad
	\gamma_5=\begin{pmatrix} -1 & 0 \\ 0 & 1 \\  \end{pmatrix}
	\ee
	where $\sigma^i$ are the standard Pauli matrices.
	
	When applying the projection operator $\mathcal{P}_+$ to the quark field, it acts as
	\be
	\mathcal{P}_+u=\mathcal{P}_+\begin{pmatrix} u_1 \\ u_2 \end{pmatrix}=\frac{1}{2}\begin{pmatrix} u_1+u_2 \\ u_1+u_2 \end{pmatrix}.
	\ee
	Here, $u_1+u_2$ represents a two-component field. Consequently, we analyze the spin structure of the correlation functions and matrix elements within a $2\times2$ spin space. 
	
	According to the convention of gamma matrices, the relationship between Euclidean and Minkowski operators at the origin is given by
	\be
	O_{\pi^{\pm,0}}^{E}(0)=O_{\pi^{\pm,0}}(0),\quad O_{p,n}^E(0)=-iO_{p,n}(0).
	\ee
	
	The action of the isospin raising and lowering operators is defined as
	\be
	I_+ d=u,\quad I_-u=d,\quad I_+\bar{u}=-\bar{d},\quad I_-\bar{d}=-\bar{u}.
	\ee
	The isospin-triplet operator for the pion, $O^{I,I_z}_\pi$, and the doublet operator for the nucleon, $O_N^{I,I_z}$, are defined as
	\ba
	&&O^{1,1}_\pi=O_{\pi^+}^E,\quad O^{1,0}_\pi=-O_{\pi^0}^E,\quad O^{1,-1}_\pi=-O_{\pi^-}^E,
	\nn\\
	&&O^{\frac{1}{2},\frac{1}{2}}_N=O_p^E,\quad O^{\frac{1}{2},-\frac{1}{2}}_N=O_n^E.
	\ea
	The isospin-triplet operators for the vector and axial-vector currents are given by
	\ba
	&&V_\mu^{1,0}=-\frac{1}{\sqrt{2}}(\bar{u}\gamma_\mu^Eu-\bar{d}\gamma_\mu^E d),
	\nn\\ 
	&&V_\mu^{0,0}=\frac{1}{\sqrt{2}}(\bar{u}\gamma_\mu^Eu+\bar{d}\gamma_\mu^E d)
	\nn\\
	&&A_\mu^{1,1}=\bar{d}\gamma_\mu^E\gamma_5^Eu,
	\nn\\
	&&A_\mu^{1,0}=-\frac{1}{\sqrt{2}}(\bar{u}\gamma_\mu^E\gamma_5^Eu-\bar{d}\gamma_\mu^E\gamma_5^Ed),
	\nn\\
	&&A_\mu^{1,-1}=-\bar{u}\gamma_\mu^E\gamma_5^Ed.
	\ea
	Using these conventions, the electromagnetic and weak currents, defined in Eq.~(\ref{eq:current}), can be expressed as
	\ba
	J_\mu^{em}(0)&=&\frac{e}{\sqrt{2}}\eta_\mu\left(-V_\mu^{1,0}(0)+\frac{1}{3}V_\mu^{0,0}(0)\right),
	\nn\\
	J_\mu^{W,A}(0)&=&-\frac{g_2}{2\sqrt{2}}\eta_\mu A_\mu^{1,1}(0),
	\nn\\
	J_\mu^{Z,A}(0)&=&\frac{g_2}{2\sqrt{2}\cos\theta_W}\eta_\mu A_\mu^{1,0}(0),
	\ea
	where $\eta_\mu=1$ for $\mu=0$ and $\eta_\mu=-i$ for $\mu=1,2,3$.

	The nucleon-pion operators in the isospin basis, $O_{N\pi}^{I,I_z}$, are given by
	\ba
	O_{N\pi}^{\frac{1}{2},\frac{1}{2}}&=&\frac{\sqrt{2}}{\sqrt{3}}O_{N}^{\frac{1}{2},-\frac{1}{2}}O_{\pi}^{1,1}-\frac{1}{\sqrt{3}}O_N^{\frac{1}{2},\frac{1}{2}}O_{\pi}^{1,0},
	\nn\\
	O_{N\pi}^{\frac{1}{2},-\frac{1}{2}}&=&\frac{1}{\sqrt{3}}O_N^{\frac{1}{2},-\frac{1}{2}}O_{\pi}^{1,0}-\frac{\sqrt{2}}{\sqrt{3}}O_N^{\frac{1}{2},\frac{1}{2}}O_{\pi}^{1,-1},
	\nn\\
	O_{N\pi}^{\frac{3}{2},\frac{3}{2}}&=&O_{N}^{\frac{1}{2},\frac{1}{2}}O_{\pi}^{1,1},
	\nn\\
	O_{N\pi}^{\frac{3}{2},\frac{1}{2}}&=&\frac{1}{\sqrt{3}}O_{N}^{\frac{1}{2},-\frac{1}{2}}O_{\pi}^{1,1}+\frac{\sqrt{2}}{\sqrt{3}}O_N^{\frac{1}{2},\frac{1}{2}}O_{\pi}^{1,0},
	\nn\\
	O_{N\pi}^{\frac{3}{2},-\frac{1}{2}}&=&\frac{\sqrt{2}}{\sqrt{3}}O_{N}^{\frac{1}{2},-\frac{1}{2}}O_{\pi}^{1,0}+\frac{1}{\sqrt{3}}O_N^{\frac{1}{2},\frac{1}{2}}O_{\pi}^{1,-1},
	\nn\\
	O_{N\pi}^{\frac{3}{2},-\frac{3}{2}}&=&O_{N}^{\frac{1}{2},-\frac{1}{2}}O_{\pi}^{1,-1}.
	\ea	
	These definitions are consistent with those given in Ref.~\cite{Wang:2023omf}.

	\subsection{State conventions}
	
	The pion decay constant, $F_\pi$, arises from the coupling between the axial vector current and the pion state, which can be expressed as
	\be
	\label{eq:decay_const}
	\langle 0|A_\mu^a(x)|\pi^b(p)\rangle=-i\,F_\pi p_\mu \delta^{ab}e^{-ip\cdot x},
	\ee
	where $A_\mu^a=\bar{\psi}\gamma_\mu\gamma_5\frac{\tau^a}{2}\psi$ is the axial vector current, and $\tau^a$  ($a=1,2,3$) are the Pauli matrices. $F_\pi$ has a value of $\sim93$ MeV.
	The pion state $|\pi^a\rangle$ satisfies the normalization condition $\langle \pi^b(p')|\pi^a(p)\rangle=2E\delta^{ab}(2\pi)^2\delta^{(3)}(\vec{p}'-\vec{p})$.
	
	The charged and neutral pion states $|\pi^{\pm,0}\rangle$ are related to $|\pi^{a}\rangle$ through
	\be
	|\pi^\pm\rangle=\frac{1}{\sqrt{2}}(|\pi^1\rangle\pm i|\pi^2\rangle),
	\quad |\pi^0\rangle=|\pi^3\rangle.
	\ee
	These states can further be related to the isospin states $|\pi^{I,I_z}\rangle$ as
	\be
	|\pi^+\rangle=|\pi^{1,1}\rangle,\quad |\pi^0\rangle=-|\pi^{1,0}\rangle,\quad |\pi^-\rangle=-|\pi^{1,-1}\rangle.
	\ee
	In a finite volume, these states are denoted as $|\pi^{I,I_z} \rangle_V$ and normalized by
	\be
	{_V}\langle \pi^{I,I_z'}|\pi^{I,I_z}\rangle_V=2E_\pi L^3\delta_{I_z,I_z'}.
	\ee
	
	The charged and neutral weak currents are related to $A_\mu^a$ through
	\ba
	&&A_\mu^+=A_\mu^1- iA_\mu^2=\bar{d}\gamma_\mu\gamma_5u=\eta_\mu A_\mu^{1,1}
	\nn\\
	&&A_\mu^-=A_\mu^1+ iA_\mu^2=\bar{u}\gamma_\mu\gamma_5d=-\eta_\mu A_\mu^{1,-1}
	\nn\\
	&& A_\mu^0=\sqrt{2}A_\mu^3=\frac{1}{\sqrt{2}}(\bar{u}\gamma_\mu\gamma_5u-\bar{d}\gamma_\mu\gamma_5d)=-\eta_\mu A_\mu^{1,0}.
	\ea
	Using this convention, we have
	\be
	\langle 0|A_\mu^{\pm,0}(x)|\pi^{\pm,0}(p)\rangle=-i\,f_\pi p_\mu e^{-ip\cdot x},
	\ee
	where $f_\pi=\sqrt{2}F_\pi\approx132$ MeV is the decay constant, typically determined from experiment or lattice QCD. 
	
	The same operator $A_\mu^{I,I_z}$ can act as either an annihilation operator or a creation operator. To distinguish these roles, 
	we use $\tilde{A}_\mu^{I,-I_z}$ to denote the $A_\mu^{I,I_z}$ operator when it functions as a creation operator. This convention also applies to the vector current.
	
	In Euclidean space, the matrix element for the axial vector current and the pion state is given by
	\be
	\langle 0|A_\mu^{I,I_z}(x)|\pi^{I,I_z'}(p^E)\rangle=-f_\pi p_\mu^E e^{-ip^E\cdot x}\,\delta_{I_z,I_z'},
	\ee
	where $\eta_\mu p_\mu^E=ip_\mu$ or equivalently $p_0^E=ip^0$ and $p_i^E=p^i$ for $i=1,2,3$. For the pion operator, we have
	\be
	\langle 0|O_\pi^{I,I_z}(x)|\pi^{I,I_z'}(p^E)\rangle=Z_\pi e^{-ip^E\cdot x}\,\delta_{I_z,I_z'},
	\ee
	with $Z_\pi$ being the corresponding overlap amplitude.
	
	The proton and neutron states can be expressed in terms of the isospin doublet $|N^{I,I_z}\rangle$ by
	\be
	|p\rangle=|N^{\frac{1}{2},\frac{1}{2}}\rangle,\quad |n\rangle=|N^{\frac{1}{2},-\frac{1}{2}}\rangle.
	\ee
	Unless otherwise specified, the spin of the state will be omitted for simplicity. 
	
	The isospin states of $N\pi$ system are defined as follows
	\ba
	&&|(N\pi)^{\frac{1}{2},\frac{1}{2}}\rangle=\frac{\sqrt{2}}{\sqrt{3}}|N^{\frac{1}{2},-\frac{1}{2}}\pi^{1,1}\rangle-\frac{1}{\sqrt{3}}|N^{\frac{1}{2},\frac{1}{2}}\pi^{1,0}\rangle,
	\nn\\
	&&|(N\pi)^{\frac{1}{2},-\frac{1}{2}}\rangle=\frac{1}{\sqrt{3}}|N^{\frac{1}{2},-\frac{1}{2}}\pi^{1,0}\rangle-\frac{\sqrt{2}}{\sqrt{3}}|N^{\frac{1}{2},\frac{1}{2}}\pi^{1,-1}\rangle,
	\nn\\
	&&|(N\pi)^{\frac{3}{2},\frac{1}{2}}\rangle=\frac{1}{\sqrt{3}}|N^{\frac{1}{2},-\frac{1}{2}}\pi^{1,1}\rangle+\frac{\sqrt{2}}{\sqrt{3}}|N^{\frac{1}{2},\frac{1}{2}}\pi^{1,0}\rangle,
	\nn\\
	&&|(N\pi)^{\frac{3}{2},-\frac{1}{2}}\rangle=\frac{\sqrt{2}}{\sqrt{3}}|N^{\frac{1}{2},-\frac{1}{2}}\pi^{1,0}\rangle+\frac{1}{\sqrt{3}}|N^{\frac{1}{2},\frac{1}{2}}\pi^{1,-1}\rangle.
	\nn\\
	\ea
	The relationship between finite-volume and infinite-volume states is given by
	\be
	|(N\pi)^{I,I_z}\rangle=\left(2\mu_{N\pi}\right)^\frac{1}{2}f_{LL}^{\frac{1}{2}}|(N\pi)^{I,I_z},\Gamma\rangle_V,
	\ee
	where the Lellouch-L\"uscher factor $f_{LL}$ is defined as~\cite{Lellouch:2000pv}
	\be
	f_{LL}=\frac{2\pi}{k^3}\left(q\frac{d\phi^\Gamma}{dq}+k\frac{d\delta}{dk}\right).
	\ee
	Here, $k$ satisfies $E_{N\pi}=\sqrt{m^2+k^2}+\sqrt{m_\pi^2+k^2}$ and $q=\frac{kL}{2\pi}$. 
	The reduced energy $\mu_{N\pi}$ is defined as $\mu_{N\pi}=\frac{EE_\pi}{E_{N\pi}}$.
	$\phi^\Gamma(q)$ is a known function associated with an irreducible representation of hypercubic symmetry, denoted by $\Gamma$. 
	In this calculation, we set $\Gamma=G_1^-$. At threshold, the large-$L$ expansion of $f_{LL}$ is given by~\cite{Feng:2018pdq}
	\be
	\label{eq:LL}
	f_{LL}=L^3\left[1+d_1\frac{a_{N\pi}}{L}+d_2\left(\frac{a_{N\pi}}{L}\right)^2+O(L^{-3})\right],
	\ee
	where $a_{N\pi}$ is the S-wave $N\pi$ scattering length, and the coefficients $d_i$ are given by
	\ba
	d_1&=&-2\frac{Z_{00}(1;0)}{\pi}=5.674\,595,
	\nn\\
	d_2&=&\frac{Z_{00}(1;0)^2+3Z_{00}(2;0)}{\pi^2}=13.075\,478.
	\ea
	The values of the zeta function $Z_{00}(s,0)$ are provided in Ref.~\cite{Luscher:1986pf}.
	The results for $a_{N\pi}$ from the two ensembles used in this calculation have already been reported in Ref.~\cite{Wang:2023omf}
	
	\subsection{Matrix element conventions}
	
	The matrix elements involving isospin states and a vector current insertion are expressed as
	\be
	\langle (N\pi)^{I'',I_z''}|\tilde{V}_\mu^{I',I_z'}(0)|N^{I,I_z}\rangle=C_{I,I_z;I',I_z'}^{I'',I_z''}\langle (N\pi)^{I''}||\tilde{V}_\mu^{I'}||N^I\rangle
	\ee
	where the Clebsch-Gordan (CG) coefficients are defined as
	\be
	C_{I,I_z;I',I_z'}^{I'',I_z''}=\langle I,I_z;I',I_z'|I'',I_z''\rangle.
	\ee
	To avoid ambiguity, we fix the $z$-components of the nucleon state and the current $\{I_{z},I_{z}'\}$ at $\{\frac{1}{2},0\}$ and define
	the following matrix elements as
	\ba
	&&\mathcal{V}_\mu^{(1)}=\langle (N\pi)^{\frac{1}{2},\frac{1}{2}}|\tilde{V}_\mu^{1,0}(0)|N^{\frac{1}{2},\frac{1}{2}}\rangle,
	\nn\\
	&&\mathcal{V}_\mu^{(2)}=\langle (N\pi)^{\frac{3}{2},\frac{1}{2}}|\tilde{V}_\mu^{1,0}(0)|N^{\frac{1}{2},\frac{1}{2}}\rangle,
	\nn\\
	&&\mathcal{V}_\mu^{(3)}=\langle (N\pi)^{\frac{1}{2},\frac{1}{2}}|\tilde{V}_\mu^{0,0}(0)|N^{\frac{1}{2},\frac{1}{2}}\rangle.
	\ea
	Using the CG coefficients, we derive the following relations
	\ba
	&&\langle (N\pi)^{\frac{1}{2},-\frac{1}{2}}|\tilde{V}_\mu^{1,0}(0)|N^{\frac{1}{2},-\frac{1}{2}}\rangle=-\mathcal{V}_\mu^{(1)},
	\nn\\
	&&\langle (N\pi)^{\frac{3}{2},-\frac{1}{2}}|\tilde{V}_\mu^{1,0}(0)|N^{\frac{1}{2},-\frac{1}{2}}\rangle=\mathcal{V}_\mu^{(2)},
	\nn\\
	&&\langle (N\pi)^{\frac{1}{2},-\frac{1}{2}}|\tilde{V}_\mu^{0,0}(0)|N^{\frac{1}{2},-\frac{1}{2}}\rangle=\mathcal{V}_\mu^{(3)}.
	\ea
	
	For physical states, the matrix elements involving the electromagnetic current are given by
	\ba
	\langle n\pi^+|J_\mu^{em}|p\rangle&=&\frac{e}{\sqrt{6}}\eta_\mu\left(-\sqrt{2}\mathcal{V}_\mu^{(1)}-\mathcal{V}_\mu^{(2)}+\frac{\sqrt{2}}{3}\mathcal{V}_\mu^{(3)}\right),
	\nn\\
	\langle p\pi^-|J_\mu^{em}|n\rangle&=&\frac{e}{\sqrt{6}}\eta_\mu\left(\sqrt{2}\mathcal{V}_\mu^{(1)}+\mathcal{V}_\mu^{(2)}+\frac{\sqrt{2}}{3}\mathcal{V}_\mu^{(3)}\right),
	\nn\\
	\langle p\pi^0|J_\mu^{em}|p\rangle&=&\frac{e}{\sqrt{6}}\eta_\mu\left(-\mathcal{V}_\mu^{(1)}+\sqrt{2}\mathcal{V}_\mu^{(2)}+\frac{1}{3}\mathcal{V}_\mu^{(3)}\right).
	\ea
	
	Matrix elements involving isospin states and an axial-vector current insertion are defined similarly
	\ba
	&&\mathcal{A}_\mu^{(1)}=\langle (N\pi)^{\frac{1}{2},\frac{1}{2}}|\tilde{A}_\mu^{1,0}(0)|N^{\frac{1}{2},\frac{1}{2}}\rangle,
	\nn\\
	&&\mathcal{A}_\mu^{(2)}=\langle (N\pi)^{\frac{3}{2},\frac{1}{2}}|\tilde{A}_\mu^{1,0}(0)|N^{\frac{1}{2},\frac{1}{2}}\rangle.
	\ea
	Using CG coefficients, we obtain
	\ba
	&&\langle (N\pi)^{\frac{1}{2},-\frac{1}{2}}|\tilde{A}_\mu^{1,-1}(0)|N^{\frac{1}{2},\frac{1}{2}}\rangle=\sqrt{2}\mathcal{A}_\mu^{(1)},
	\nn\\
	&&\langle (N\pi)^{\frac{3}{2},-\frac{1}{2}}|\tilde{A}_\mu^{1,-1}(0)|N^{\frac{1}{2},\frac{1}{2}}\rangle=\frac{1}{\sqrt{2}}\mathcal{A}_\mu^{(2)},
	\ea
	
	For physical states, the matrix elements involving the weak current are 
	\ba
	\langle p\pi^-|J_\mu^{W,A}|p\rangle&=&-\frac{g_2}{2\sqrt{6}}\eta_\mu\left(2\mathcal{A}_\mu^{(1)}-\frac{1}{\sqrt{2}}\mathcal{A}_\mu^{(2)}\right),
	\nn\\
	\langle n\pi^0|J_\mu^{W,A}|p\rangle&=&-\frac{g_2}{2\sqrt{6}}\eta_\mu\left(-\sqrt{2}\mathcal{A}_\mu^{(1)}-\mathcal{A}_\mu^{(2)}\right),
	\nn\\
	\langle p\pi^0|J_\mu^{Z,A}|p\rangle&=&\frac{g_2}{2\sqrt{6}\cos\theta_W}\eta_\mu\left(\mathcal{A}_\mu^{(1)}-\sqrt{2}\mathcal{A}_\mu^{(2)}\right),
	\nn\\
	\langle n\pi^+|J_\mu^{Z,A}|p\rangle&=&\frac{g_2}{2\sqrt{6}\cos\theta_W}\eta_\mu\left(\sqrt{2}\mathcal{A}_\mu^{(1)}+\mathcal{A}_\mu^{(2)}\right).
	\nn\\
	\ea
	
	The multipole amplitudes in the physical and isospin bases can be related in the same manner. We summarize the relationship
	using $L_{0+}$ and $L_{0+}^{(W)}$ as examples
 	\ba
	&&L_{0+}(\gamma^* p\to n\pi^+)=\frac{e}{\sqrt{6}}\left(-\sqrt{2}L_{0+}^{(1)}-L_{0+}^{(2)}+\frac{\sqrt{2}}{3}L_{0+}^{(3)}\right),
	\nn\\
	&&L_{0+}(\gamma^* n\to p\pi^-)=\frac{e}{\sqrt{6}}\left(\sqrt{2}L_{0+}^{(1)}+L_{0+}^{(2)}+\frac{\sqrt{2}}{3}L_{0+}^{(3)}\right),
	\nn\\
	&&L_{0+}(\gamma^* p\to p\pi^0)=\frac{e}{\sqrt{6}}\left(-L_{0+}^{(1)}+\sqrt{2}L_{0+}^{(2)}+\frac{1}{3}L_{0+}^{(3)}\right),
	\nn\\
	&&L_{0+}^{(W)}(W^{-*} p\to p\pi^-)=-\frac{g_2}{2\sqrt{6}}\left(2L_{0+}^{(W),(1)}-\frac{1}{\sqrt{2}}L_{0+}^{(W),(2)}\right),
	\nn\\
	&&L_{0+}^{(W)}(W^{-*} p\to n\pi^0)=-\frac{g_2}{2\sqrt{6}}\left(-\sqrt{2}L_{0+}^{(W),(1)}-L_{0+}^{(W),(2)}\right),
	\nn\\
	&&L_{0+}^{(Z)}(Z^{0*} p\to p\pi^0)=\frac{g_2}{2\sqrt{6}\cos\theta_W}\left(L_{0+}^{(W),(1)}-\sqrt{2}L_{0+}^{(W),(2)}\right),
	\nn\\
	&&L_{0+}^{(Z)}(Z^{0*} p\to n\pi^+)=\frac{g_2}{2\sqrt{6}\cos\theta_W}\left(\sqrt{2}L_{0+}^{(W),(1)}+L_{0+}^{(W),(2)}\right).
	\nn\\
	\ea
	
	\subsection{Spin projection for extracting the multipole amplitudes}
	
	We define the $2\times2$ vector matrix $\vec{\Sigma}$  as
	\be
	\vec{\Sigma}_{s',s}=\xi_{s'}^\dagger\vec{\sigma}\xi_{s}.
	\ee
	These matrices satisfy the relation
	\be
	\operatorname{Tr}[\vec{\Sigma} \cdot \vec{\Sigma}]=\sum_{s,s'}\vec{\Sigma}_{s',s}\vec{\Sigma}_{s,s'}=6.
	\ee
	Based on rotation symmetry, a natural way to extract the multipole amplitudes is to perform the spin projection as follows
	\ba
	\label{eq:spin_projection}
	&&\frac{1}{N_R}\sum_{\hat{R}\in O_h} \frac{1}{2}\operatorname{Tr}\left[(\hat{k}\cdot\vec{\Sigma})\hat{k}\cdot \vec{\mathcal{J}}^{em}\right]\Big|_{\vec{k}=\hat{R}\vec{l}}=\alpha_m\,L_{0+},
	\nn\\
	&&\frac{1}{N_R}\sum_{\hat{R}\in O_h} \frac{1}{2}\operatorname{Tr}\left[\vec{\Sigma}\cdot \vec{\mathcal{J}}^{em}\right]\Big|_{\vec{k}=\hat{R}\vec{l}}=\alpha_m\left(2E_{0+}+L_{0+}\right),
	\nn\\
	&&\frac{1}{N_R}\sum_{\hat{R}\in O_h} \frac{1}{2}\operatorname{Tr}\left[\frac{k_0}{|\vec{k}|}(\hat{k}\cdot\vec{\Sigma})\mathcal{J}_0^{em}\right]\Big|_{\vec{k}=\hat{R}\vec{l}}=\alpha_m\,L_{0+},
	\nn\\
	&&\frac{1}{N_R}\sum_{\hat{R}\in O_h} \frac{1}{2}\operatorname{Tr}\left[\mathcal{J}_0^{W,A}\right]\Big|_{\vec{k}=\hat{R}\vec{l}}=\alpha_m\left(L_{0+}^{(W)}+\frac{k_0}{m}H_{0+}\right),
	\nn\\
	&&\frac{1}{N_R}\sum_{\hat{R}\in O_h} \frac{1}{2}\operatorname{Tr}\left[\frac{m}{|\vec{k}|}\hat{k}\cdot\vec{\mathcal{J}}^{W,A}\right]\Big|_{\vec{k}=\hat{R}\vec{l}}=\alpha_m\,H_{0+},
	\nn\\
	&&\frac{1}{N_R}\sum_{\hat{R}\in O_h} \frac{1}{2}\operatorname{Tr}\left[i(\vec\Sigma\times\hat{k})\cdot\vec{\mathcal{J}}^{W,A}\right]\Big|_{\vec{k}=\hat{R}\vec{l}}=2\alpha_m\,M_{0+},
	\ea
	where $\vec{l}$ is a typical lattice momentum. 
	
	Combining Eq.~(\ref{eq:spin_projection}) and Eq.~(\ref{eq:four_point_func}) yields the formula for extracting the multipole amplitude from the correlation functions, 
	as given in Eq.~(\ref{eq:t_dep_2}), up to the finite-volume corrections.

	\subsection{Operator construction for GEVP}
	
	First, we consider the correlation function of the general form
	\be
	C(\Phi,t)=\langle O_{N(\vec{q})\pi(\vec{q}_1)\cdots \pi(\vec{q}_m)}(t)\bar{O}_{N(\vec{p})\pi(\vec{p}_1)\cdots \pi(\vec{p}_n)}(0)\rangle,
	\ee
	where the composite operator $O_{N(\vec{q})\pi(\vec{q}_1)\cdots \pi(\vec{q}_m)}$ consists of a nucleon operator and $m$ pion operators, while
	$\bar{O}_{N(\vec{p})\pi(\vec{p}_1)\cdots \pi(\vec{p}_n)}$ contains a nucleon operator and $n$ pion operators.
	In total, the momenta involved are given by
	\be
	\Phi=\{\vec{q},\vec{q}_1,\cdots,\vec{q}_m,\vec{p},\vec{p}_1,\cdots,\vec{p}_n\},
	\ee
   	where there are $m+n+2$ momenta, which satisfy the constraint
	\be
	\vec{q}+\sum_{i=1}^m\vec{q}_m+\vec{p}+\sum_{i=1}^n\vec{p}_n=0.
	\ee
	
	The general structure of $C(\Phi,t)$ can be expressed as 
	\be
	\label{eq:decomposition}
	C(\Phi,t)=C^{(0)}(\Phi,t) +C^{(i)}(\Phi,t)\sigma^i.
	\ee
	Under cubic symmetry rotations, the nucleon operator transforms as
	\be
	O_{N(\vec{q})} \to \Lambda_{\frac{1}{2}}O_{N(R^{-1}\vec{q})},\quad \bar{O}_{N(\vec{p})} \to \bar{O}_{N(R^{-1}\vec{p})}\Lambda_{\frac{1}{2}}^{-1}
	\ee
	where $\Lambda_\frac{1}{2}$ is a $2\times2$ matrix satisfying
	\be
	\Lambda_\frac{1}{2}^{-1}\sigma^i\Lambda_\frac{1}{2}=R_{ij}\sigma^j.
	\ee
	Accordingly, the correlation function transforms as
	\be
	C(\Phi,t)\to \Lambda_{\frac{1}{2}} C(R^{-1}\Phi,t)\Lambda_{\frac{1}{2}}^{-1},
	\ee
	leading to the conditions
	\ba
	\label{eq:rotation}
	&&C^{(0)}(R^{-1}\Phi,t)=C^{(0)}(\Phi,t),
	\nn\\
	&&C^{(i)}(R^{-1}\Phi,t)\sigma^i=R_{ij}C^{(i)}(\Phi,t)\sigma^j.
	\ea
	Under parity transformation, the correlation function satisfies
	\be
	C(\Phi,t)\to (-1)^{m+n}C(-\Phi,t),
	\ee
	which imposes the condition
	\be
	\label{eq:parity}
	C^{(i)}(\Phi)=(-1)^{m+n}C^{(i)}(-\Phi),\quad i=0,1,2,3.
	\ee
	
	Using Eqs.~(\ref{eq:rotation}) and (\ref{eq:parity}), we obtain
	\ba
	&&\langle O_{N(-\vec{p})}\bar{O}_{N(\vec{p})}\rangle =f_1(\vec{p}^2,t),
	\nn\\
	&&\langle O_{N(-\vec{p})}\bar{O}_{N(\vec{p})\pi(\vec{0})}\rangle = \hat{p}\cdot\vec{\sigma} f_2(\vec{p}^2,t),
	\nn\\
	&&\langle O_{N(-\vec{p})}\bar{O}_{N(\vec{0})\pi(\vec{p})}\rangle =\hat{p}\cdot\vec{\sigma}f_3(\vec{p}^2,t),
	\nn\\
	&&\langle O_{N(-\vec{p})\pi(\vec{0})}\bar{O}_{N(\vec{p})\pi(\vec{0})}\rangle = f_4(\vec{p}^2,t),
	\nn\\
	&&\langle O_{N(-\vec{p})\pi(\vec{0})}\bar{O}_{N(\vec{0})\pi(\vec{p})}\rangle = f_5(\vec{p}^2,t),
	\nn\\
	&&\langle O_{N(\vec{0})\pi(-\vec{p})}\bar{O}_{N(\vec{0})\pi(\vec{p})}\rangle = f_6(\vec{p}^2,t),
	\ea
	where the functions $f_i(\vec{p}^2,t)$ have a trivial spin structure, meaning they are proportional to the $2\times2$ identity matrix in spin space.
	By multiplying the operators $O_{N(\vec{p})\pi(\vec{0})}$ and $O_{N(\vec{0})\pi(\vec{\vec{p}})}$ by $\hat{p}\cdot\vec{\sigma}$ and defining the operators as given in Eq.~(\ref{eq:N_GEVP}),
	we demonstrate that each correlation function 
	$C_N^{n,m}(t)=\langle O_{N}^{(n)}(t)\bar{O}_{N}^{(m)}(0)\rangle$ has a trivial spin structure.
	The trace of the correlation function, denoted as $\mathcal{M}_N^{n,m}(t)$ in Eq.~(\ref{eq:correlation_matrix_N}), is given by
	\ba
	\label{eq:M_state_insertion}
	\mathcal{M}_N^{n,m}(t)&=&\frac{1}{2}\left\{\left[C_{N}^{n,m}(t)\right]_{11}+\left[C_{N}^{n,m}(t)\right]_{22}\right\}
	\nn\\
	&=&\sum_k\langle 0|O_{N,1}^{(n)}(0)|k\rangle \frac{e^{-E_kt}}{2E_k}\langle k|\bar{O}_{N,1}^{(m)}(0)|0\rangle,
	\nn\\
	\ea
	where $|k\rangle$ represents the eigenstate of the QCD Hamiltonian in a finite volume, and $O_{N,\alpha}^{(n)}$ denotes the $\alpha_{\mathrm{th}}$ component of the operator $O_N^{(n)}$.
	Through Eq.~(\ref{eq:M_state_insertion}), we establish that the correlation function matrix, whose elements are given by $\mathcal{M}_N^{n,m}(t)$,
	is well-suited for a GEVP analysis.
	
    	For nucleon-pion operator in the $G_1^-$ representation, we introduce the correlation function
  	\be
    	C_{N\pi}^{n,m}(t)=\langle O_{N\pi}^{(n)}(t)O_{N\pi}^{(m)}(0)\rangle.
   	\ee
	The operators $O_{N\pi}^{(n)}$ and $O_{N\pi}^{(m)}$ are associated with the momenta $\vec{p}$ and $\vec{q}$, respectively.
	Under rotational symmetry, $C_{N\pi}^{n,m}(t)$ can, in principle, contain terms proportional to $1_{2\times2}$, $\vec{p}\cdot\vec{\sigma}$,  $\vec{q}\cdot\vec{\sigma}$ and $\epsilon^{ijk}p^jq^k\sigma^i$.
	However, parity symmetry forbids the $\vec{p}\cdot\vec{\sigma}$ and $\vec{q}\cdot\vec{\sigma}$ terms.  Furthermore, in the $G_1^-$ representation, 
	the symmetry $\vec{p}\to-\vec{p}$ holds, eliminating the $\epsilon^{ijk}p^jq^k\sigma^i$ term as well, leaving only the $1_{2\times2}$ contribution.
	As a result, we obtain
	\be
	C_{N\pi}^{n,m}(t)=f^{n,m}(\vec{p}^2,\vec{p}\cdot\vec{q},\vec{q}^2,t),
	\ee
	where the functions $f^{n,m}$ exhibit a trivial spin structure. Following a similar procedure as described earlier, we can demonstrate that the trace of $C_{N\pi}^{n,m}(t)$, 
	denoted as $\mathcal{M}_{N\pi}^{n,m}(t)$ in Eq.~(\ref{eq:correlation_matrix_Npi}),
	is also suitable for a GEVP analysis.

	\begin{figure*}[htb]
	\centering
	\includegraphics[width=0.75\linewidth,angle=0]{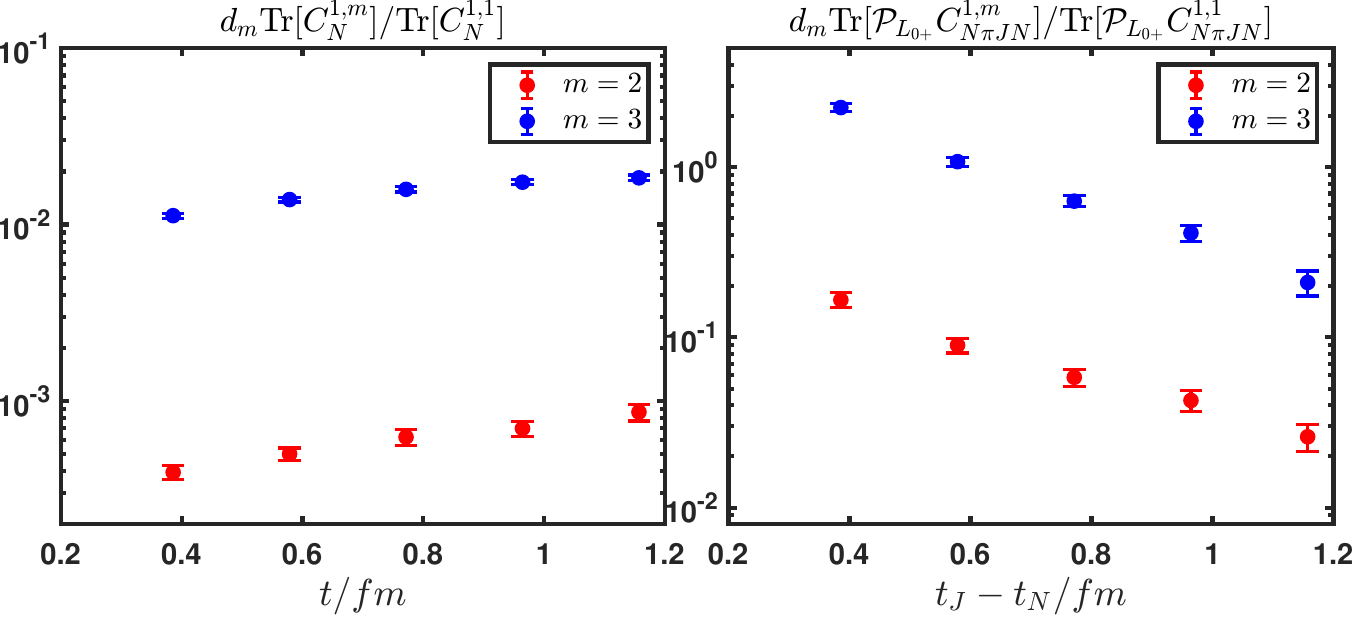}
	\caption{Effect of GEVP optimization. The left panel shows the ratio $d_m\operatorname{Tr}[C_{N}^{1,m}(t)]/\operatorname{Tr}[C_{N}^{1,1}(t)]$ as a function of $t$ for $m=2,3$.
	The right panel presents the ratio $d_m\operatorname{Tr}[\mathcal{P}_{L_{0+}} C_{N\pi JN}^{1,m}]/\operatorname{Tr}[\mathcal{P}_{L_{0+}} C_{N\pi JN}^{1,1}]$ as a function
	of $t_J-t_N$, with the $t_{N\pi}-t_J$ dependence removed through fitting. The results are shown for the 24D ensemble, focusing on the isospin channel with $I_{N\pi}=\frac{1}{2}$
	and $I_J=1$ as an example.}
	\label{fig:correction}
        \end{figure*}

	As nucleon operators with nonzero momentum are widely used in lattice calculations, it is important to examine how the GEVP procedure 
	affects the single-nucleon two-point correlation function. To assess the impact of the GEVP-optimized operator defined in Eq.~(\ref{eq:nucleon_GEVP}), we introduce the
	ratio $d_m\operatorname{Tr}[C_{N}^{1,m}(t)]/\operatorname{Tr}[C_{N}^{1,1}(t)]$, where the denominator, $\operatorname{Tr}[C_{N}^{1,1}(t)]$,  
	represents the correlation function without GEVP, while the numerator, $d_m\operatorname{Tr}[C_{N}^{1,m}(t)]$, captures the primary effect of GEVP optimization. As shown in Fig.~\ref{fig:correction},
	using the 24D ensemble as an example,
	GEVP causes only a minor $\sim$1\% change in the nucleon two-point function. However, the effect of GEVP is much more pronounced for 
	the correlation function $C_{N\pi JN}^{n,m}(t_{N\pi},t_J,t_N)=\langle O_{N\pi}^{(n)}(t_{N\pi})\tilde{J}_\mu(t_J)O_N^{(m)}(t_N)\rangle$. By applying the projection operator $\mathcal{P}_{L_{0+}}$ and focusing on the isospin channel with $I_{N\pi}=\frac{1}{2}$
	and $I_J=1$, we plot the ratio $d_m\operatorname{Tr}[\mathcal{P}_{L_{0+}} C_{N\pi JN}^{1,m}]/\operatorname{Tr}[\mathcal{P}_{L_{0+}} C_{N\pi JN}^{1,1}]$ as a function
	of $t_J-t_N$, after removing the $t_{N\pi}-t_J$ dependence through fitting. It is evident that GEVP has a significant impact on $C_{N\pi JN}^{1,1}$, at the level of O(100\%).
	This enhancement arises because $C_{N\pi JN}^{1,m}$ is substantially larger than $C_{N\pi JN}^{1,1}$. For example, taking $m=3$
	and using a factorization approximation,
	$C_{N\pi JN}^{1,3}\sim \langle O_{N(\vec{0})\pi(\vec{0})}(t_{N\pi})\tilde{J}_\mu(t_J)\bar{O}_{N(\vec{0})\pi(\vec{p})}(t_N)\rangle$ receives a dominant contribution from
	\ba
	\label{eq:factorization_form}
	\langle O_{N(\vec{0})}(t_{N\pi})\bar{O}_{N(\vec{0})}(t_N)\rangle \langle O_{\pi(\vec{0})}(t_{N\pi})\tilde{J}_\mu(t_J)\bar{O}_{\pi(\vec{p})}(t_N)\rangle.
	\nn\\
	\ea
	This term is significantly larger than $C_{N\pi JN}^{1,1}=\langle O_{N(\vec{0})\pi(\vec{0})}(t_{N\pi})\tilde{J}_\mu(t_J)\bar{O}_{N(\vec{p})}(t_N) \rangle$.
	The latter requires the operator $\tilde{J}_\mu$  to create a pion from the nucleon state, which is highly suppressed due to the substantial momentum redistribution 
	needed for a direct transition from a single-nucleon state with momentum $\vec{p}$ to a nucleon-pion system at rest. 
	In contrast, in the factorized form in Eq.~(\ref{eq:factorization_form}), the nucleon state with zero momentum is already present, avoiding this suppression.	

	\subsection{Multipole amplitudes in the isospin basis}

	\begin{figure*}[htb]
	\centering
	\captionsetup[subfigure]{font=small}
	\begin{tabular}{cc} 
	\begin{subfigure}{0.49\textwidth}
	\centering
	\includegraphics[width=\linewidth,angle=0]{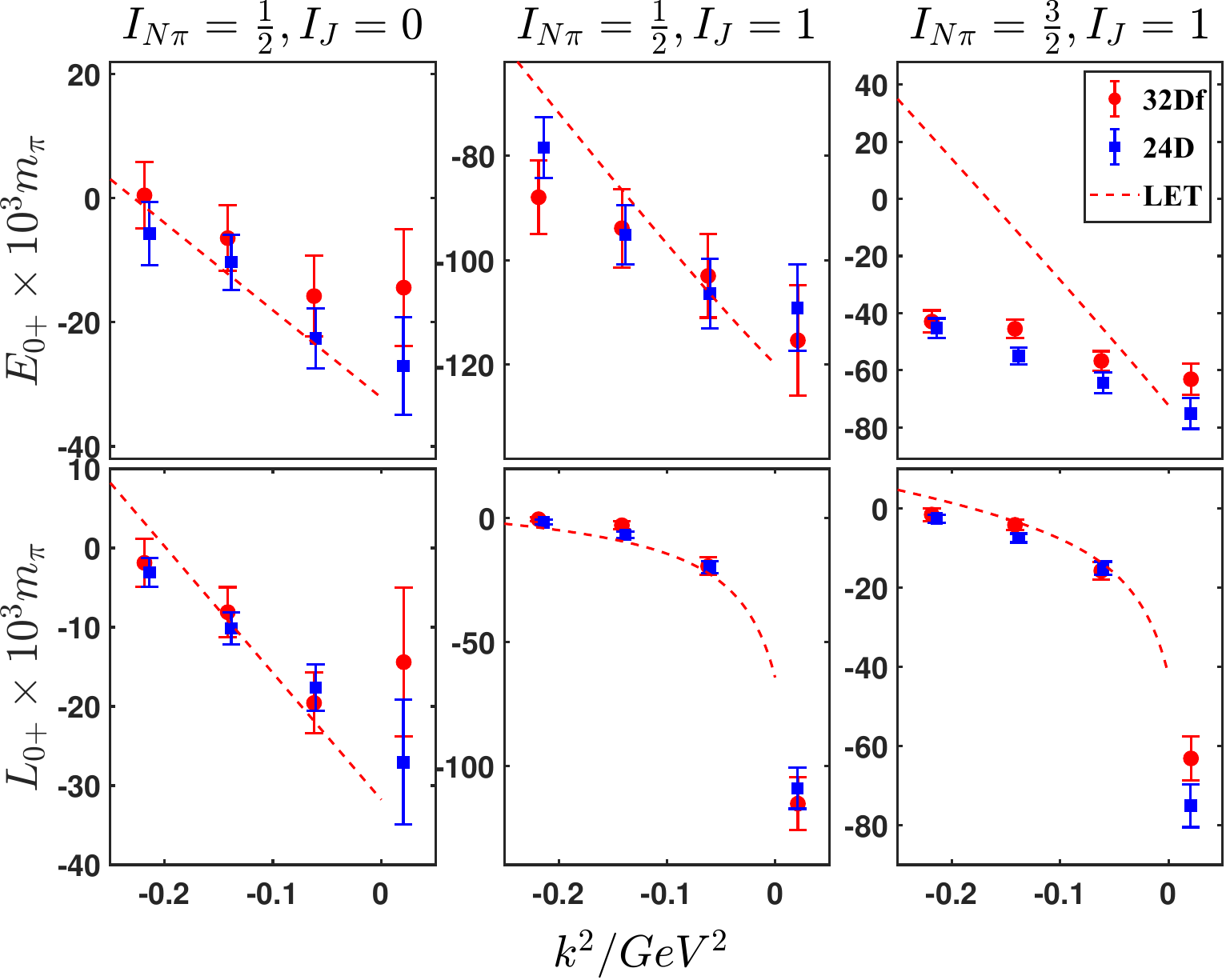}
	\caption{$L_{0+}$ and $E_{0+}$ from the electromagnetic process}
        \end{subfigure} &
        	\begin{subfigure}{0.49\textwidth}
	\centering
	\includegraphics[width=\linewidth,angle=0]{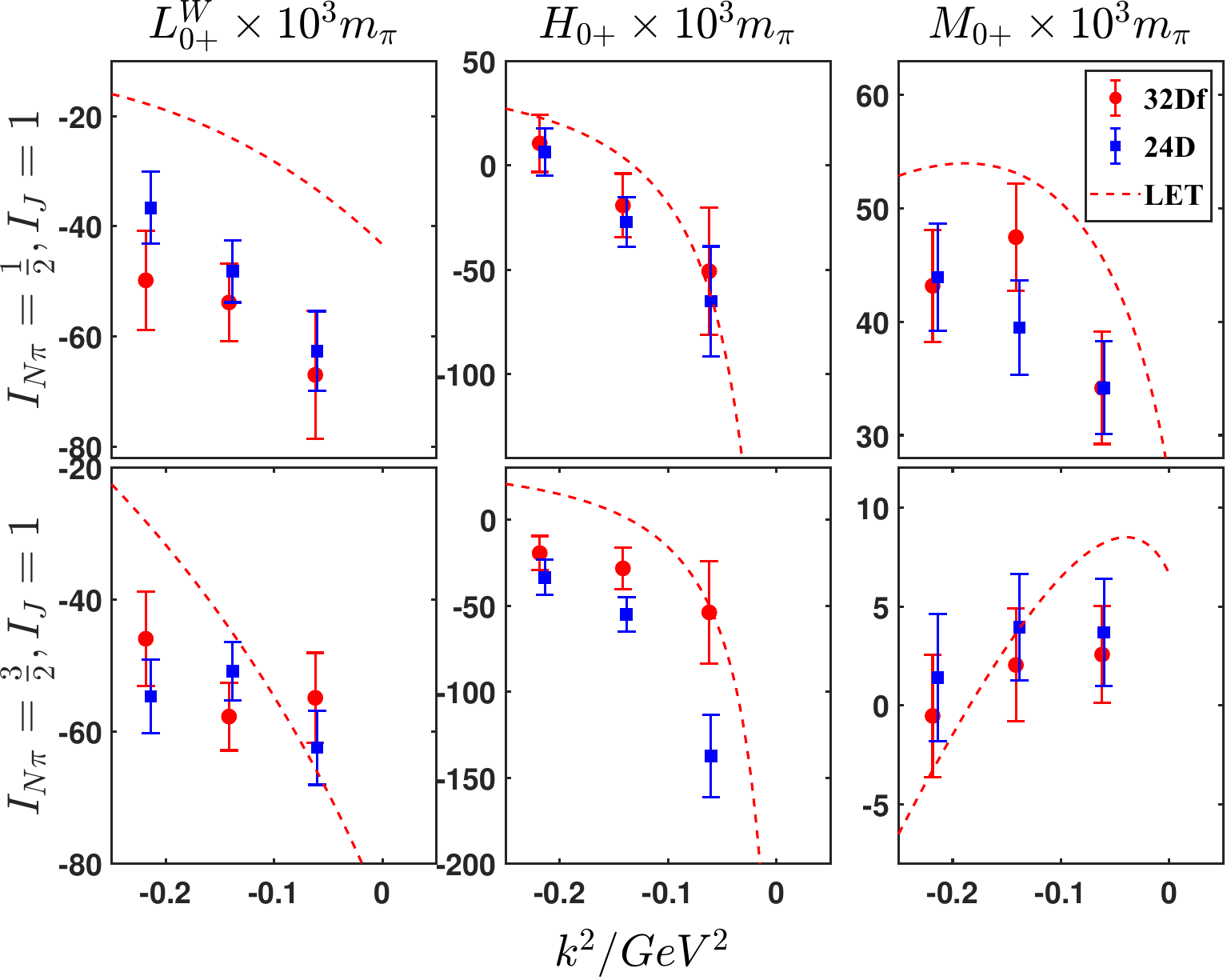}
	\caption{$L_{0+}^{(W)}$, $H_{0+}$ and $M_{0+}$ from the weak process}
         \end{subfigure}
	\end{tabular}
	\caption{Comparison of lattice results for five multipole amplitudes with LET predictions.}
	\label{fig:isospin_basis}
	\end{figure*}
	
	In Fig.~\ref{fig:isospin_basis}, we present the momentum dependence of the multipole amplitudes in the isospin basis.
	The left panel shows $L_{0+}$ and $E_{0+}$ from the electromagnetic pion production, while the right panel displays
	$L_{0+}^{(W)}$, $H_{0+}$ and $M_{0+}$ from the weak pion production. 
	The determination of these multipole amplitudes is discussed in the next subsection.
	We compare the lattice results with predictions from LETs~\cite{Bernard:1993bq,Bernard:1993xh}. 
	These LET predictions include the $O(\mu^2,\nu)$ corrections but omit higher-order contributions, where $\mu=\frac{M_\pi}{m}$
	is the pion-to-nucleon mass ratio, and $\nu=\frac{k^2}{m^2}$ represents the ratio of the squared momentum transfer carried by the currents to the squared nucleon mass.
	The lattice data exhibit a trend similar to LET predictions but show noticeable deviations, highlighting the importance of incorporating higher-order corrections in LETs. 
	On the lattice side, further dedicated efforts are needed to obtain results with a complete error budget. Given the rapid advancements in the field, 
	such improvements can be anticipated in the near future.
	It would be interesting to examine photoproduction at $k^2=0$, where much more precise ChPT predictions and experimental data are available.
	Achieving an extrapolation to $k^2=0$ requires a careful interplay between lattice QCD and ChPT.

	\subsection{Extraction of multipole amplitudes}
	
	In this section, we present figures illustrating the fitting procedure used to extract the multipole amplitudes. The dataset includes various multipole amplitudes, isospin channels, and momentum modes, 
	resulting in approximately 40 figures in total. Since the time dependence is similar across different momentum modes, we display only the $(0,0,1)$ mode as a representative example to avoid redundancy.
		
\begin{figure*}[htb]
	\centering
	\captionsetup[subfigure]{font=footnotesize}
	\begin{tabular}{cc} 
	\begin{subfigure}{0.48\textwidth}
	\centering
	\includegraphics[width=\linewidth,angle=0]{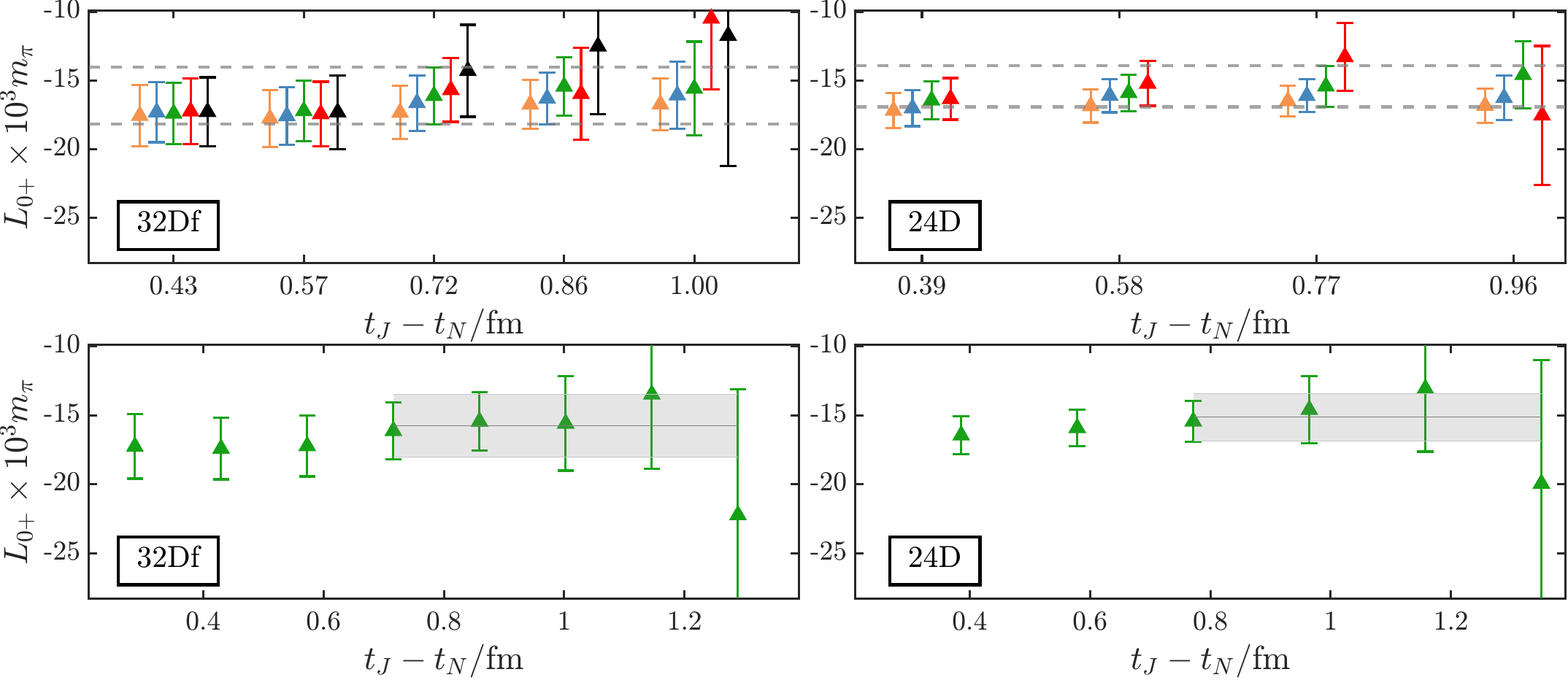}
	\caption{$L_{0+}$: $(I_{N\pi},I_J,I_N)=(\frac{3}{2},1,\frac{1}{2})$}
        \end{subfigure} &
        	\begin{subfigure}{0.48\textwidth}
	\centering
	\includegraphics[width=\linewidth,angle=0]{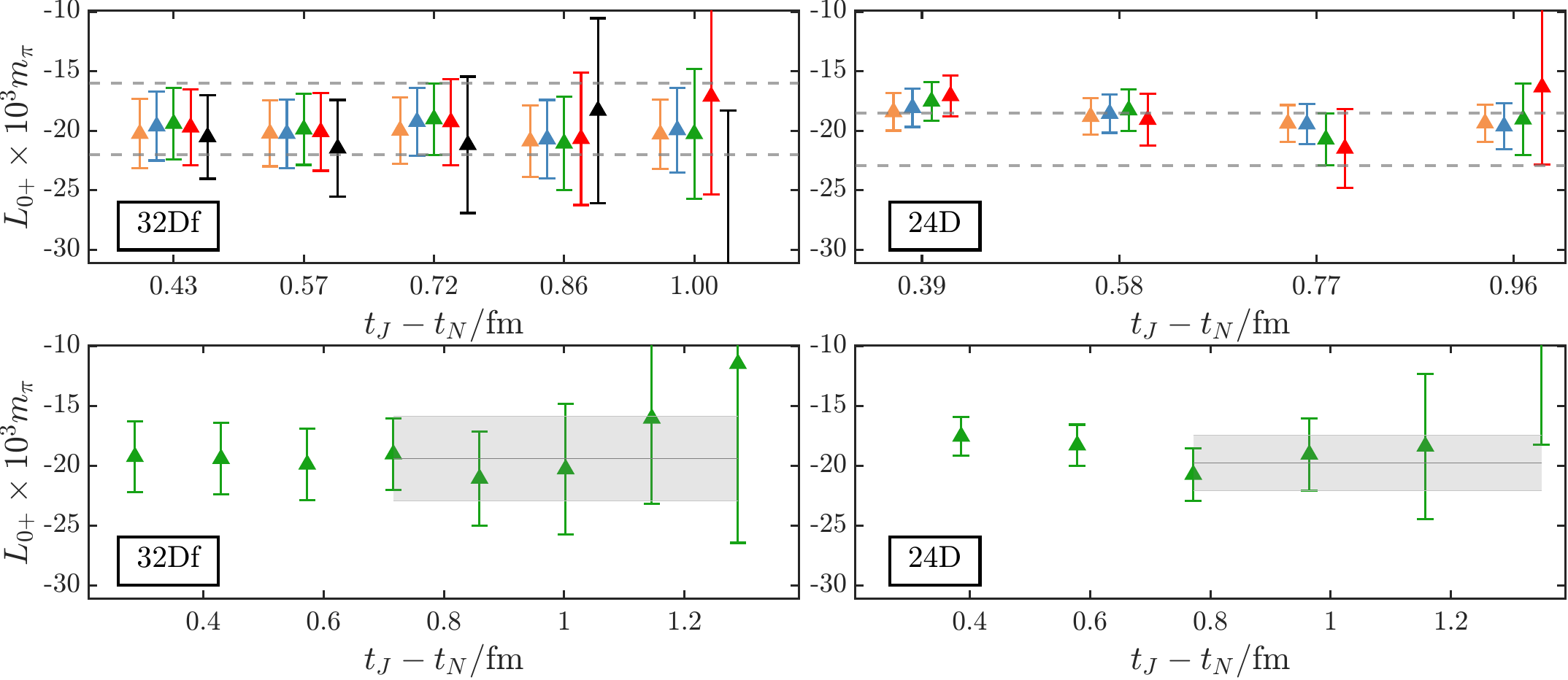}
	\caption{$L_{0+}$: $(I_{N\pi},I_J,I_N)=(\frac{1}{2},1,\frac{1}{2})$}
         \end{subfigure}\\
       \begin{subfigure}{0.48\textwidth}
	\centering
	\includegraphics[width=\linewidth,angle=0]{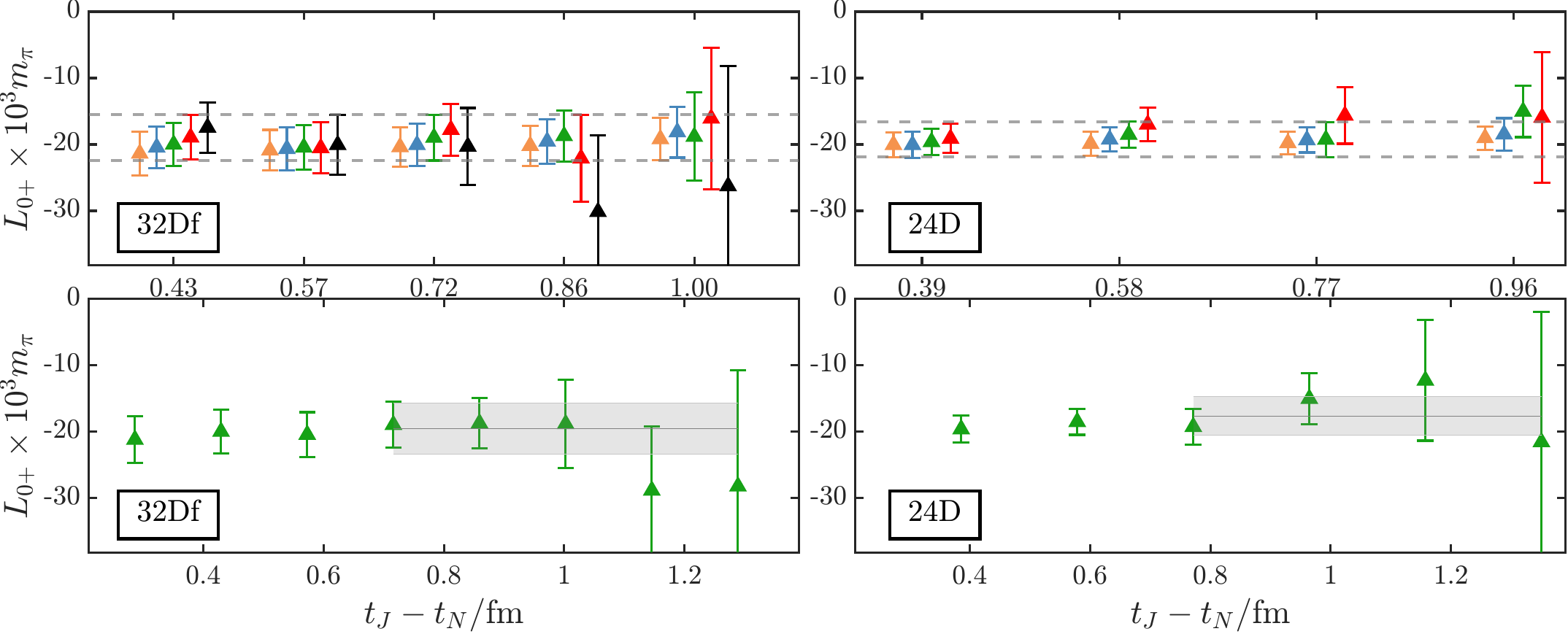}
	\caption{$L_{0+}$: $(I_{N\pi},I_J,I_N)=(\frac{1}{2},0,\frac{1}{2})$}
       \end{subfigure} &
       \begin{subfigure}{0.48\textwidth}
	\centering
	\includegraphics[width=\linewidth,angle=0]{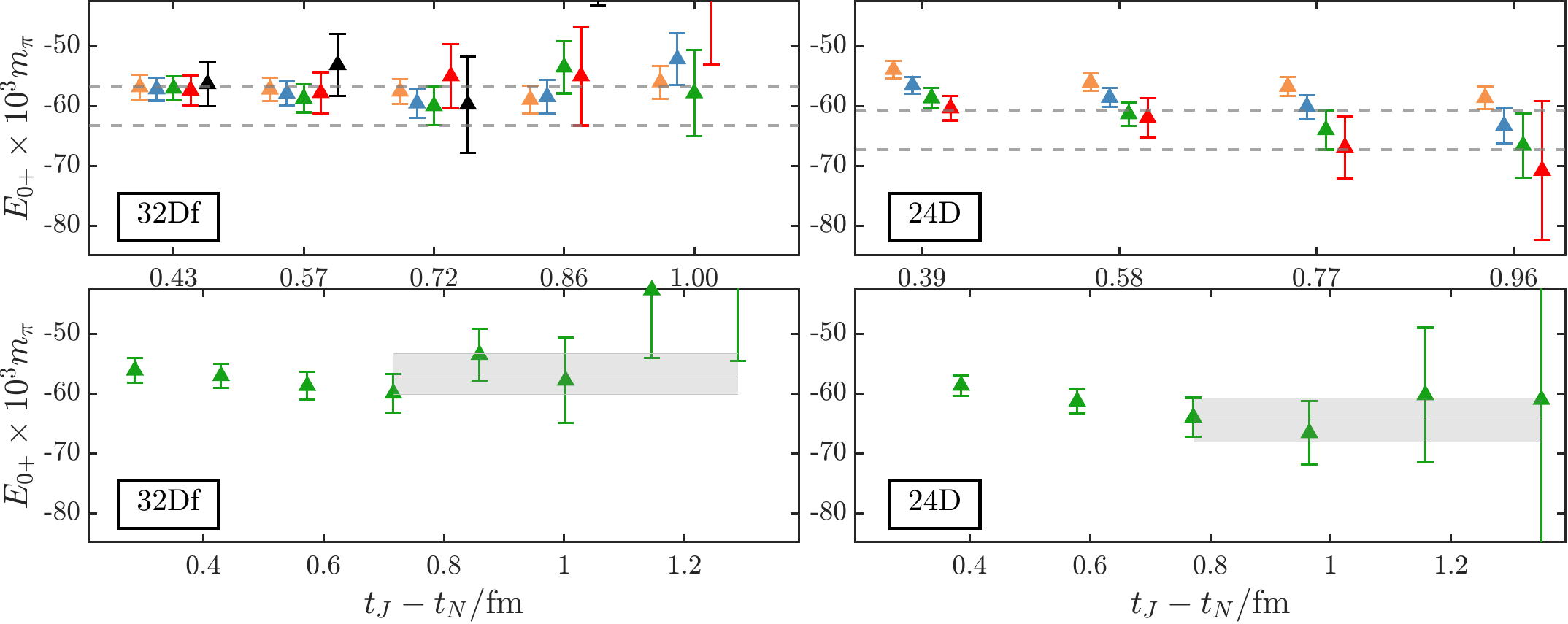}
	\caption{$E_{0+}$: $(I_{N\pi},I_J,I_N)=(\frac{3}{2},1,\frac{1}{2})$}
       \end{subfigure}\\      
	\begin{subfigure}{0.48\textwidth}
	\centering
	\includegraphics[width=\linewidth,angle=0]{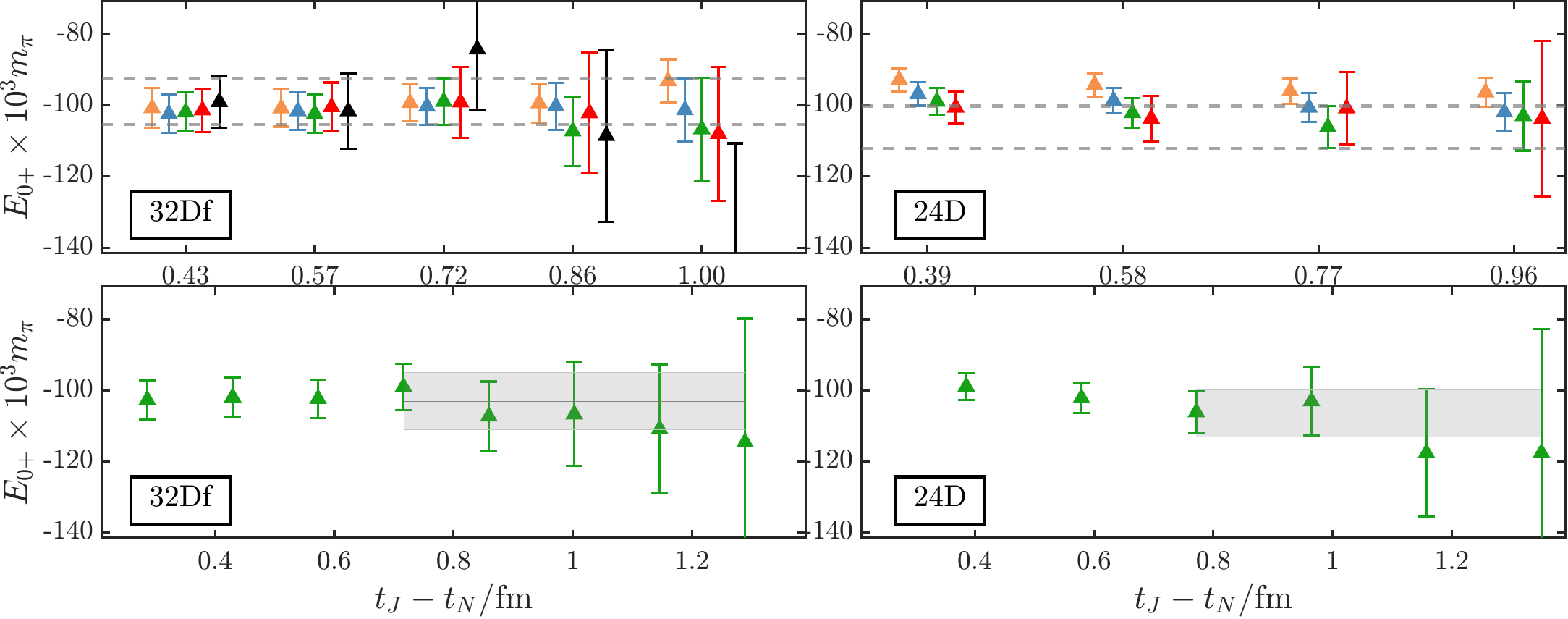}
	\caption{$E_{0+}$: $(I_{N\pi},I_J,I_N)=(\frac{1}{2},1,\frac{1}{2})$}
        \end{subfigure} &
        	\begin{subfigure}{0.48\textwidth}
	\centering
	\includegraphics[width=\linewidth,angle=0]{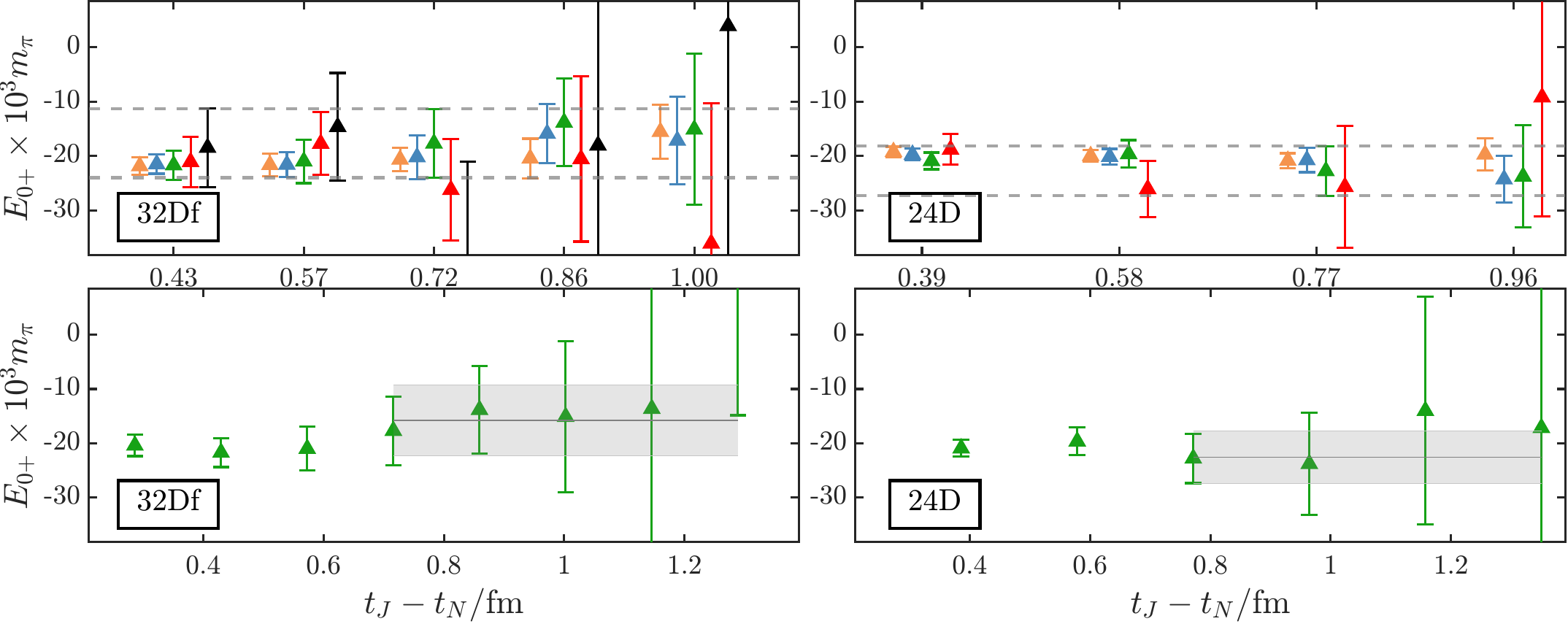}
	\caption{$E_{0+}$: $(I_{N\pi},I_J,I_N)=(\frac{1}{2},0,\frac{1}{2})$}
         \end{subfigure}\\
       \begin{subfigure}{0.48\textwidth}
	\centering
	\includegraphics[width=\linewidth,angle=0]{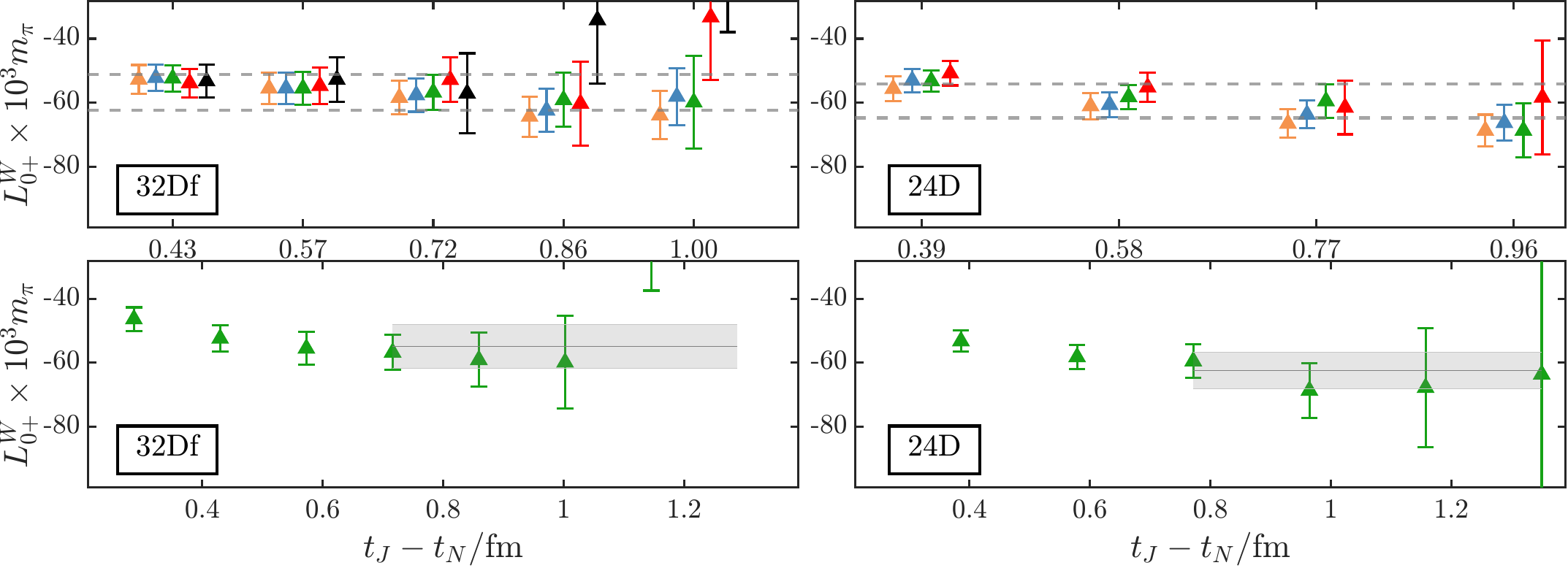}
	\caption{$L_{0+}^{(W)}$: $(I_{N\pi},I_J,I_N)=(\frac{3}{2},1,\frac{1}{2})$}
       \end{subfigure} &
       \begin{subfigure}{0.48\textwidth}
	\centering
	\includegraphics[width=\linewidth,angle=0]{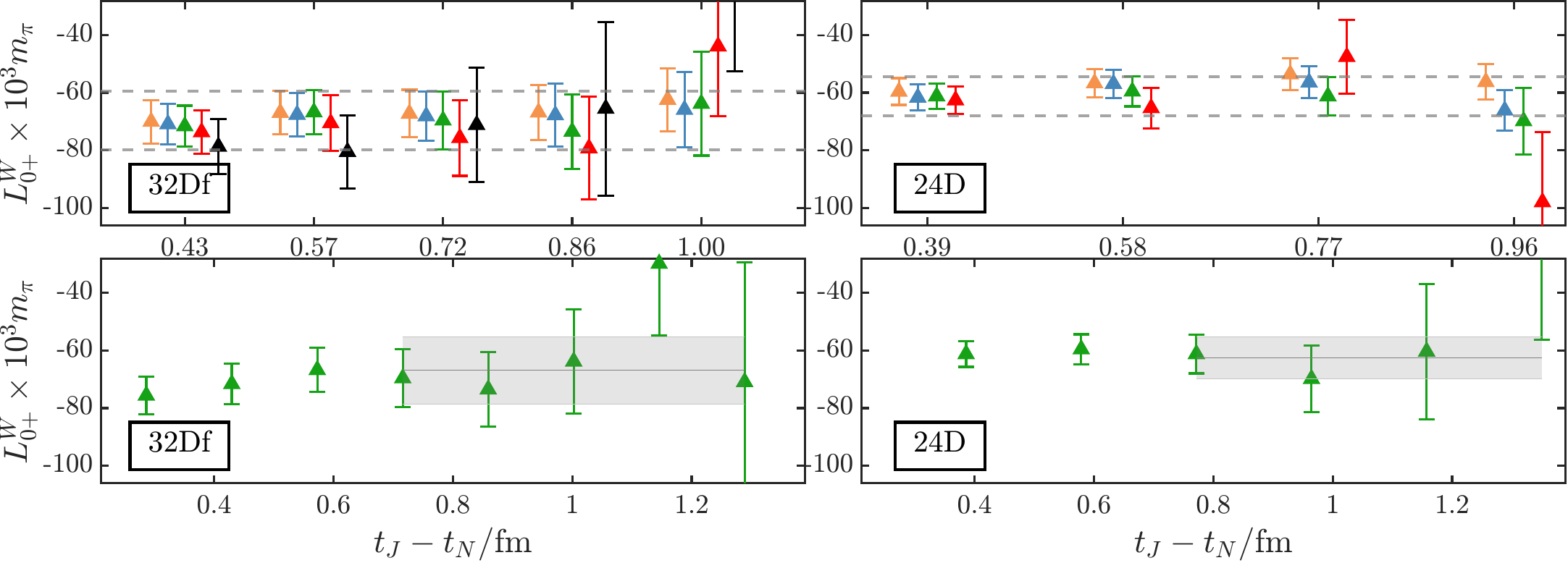}
	\caption{$L_{0+}^{(W)}$: $(I_{N\pi},I_J,I_N)=(\frac{1}{2},1,\frac{1}{2})$}
       \end{subfigure}\\      
	\begin{subfigure}{0.48\textwidth}
	\centering
	\includegraphics[width=\linewidth,angle=0]{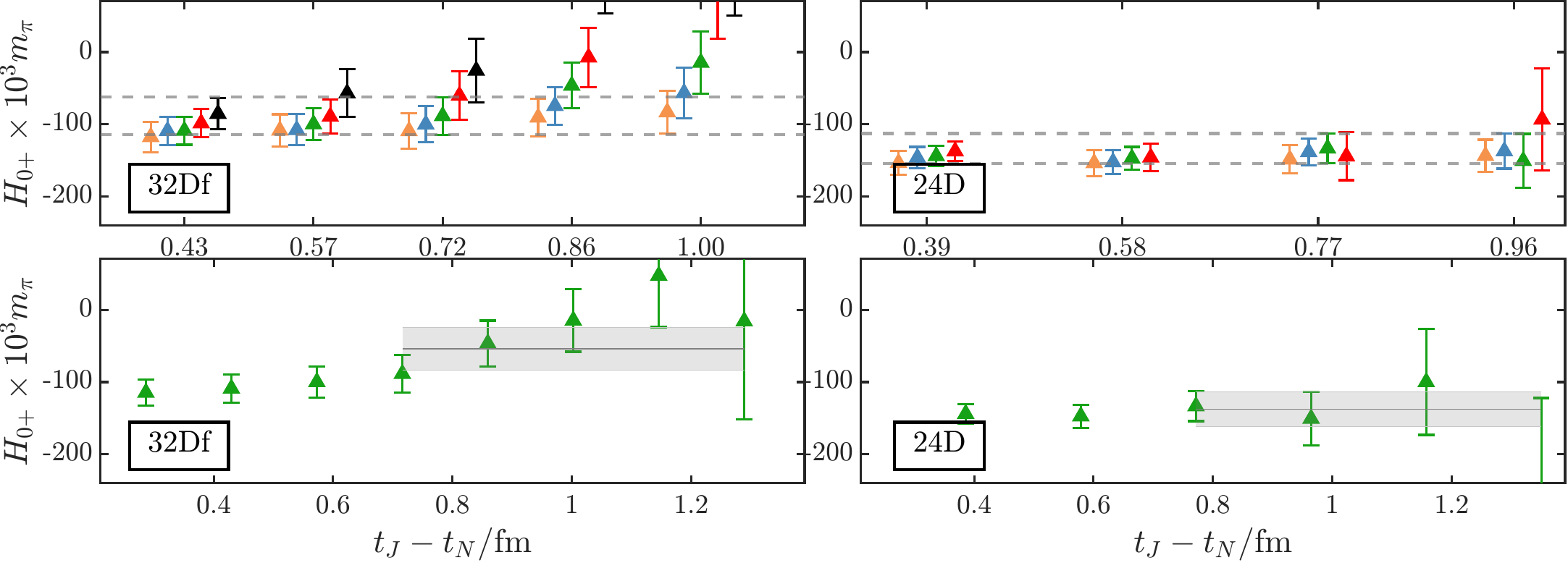}
	\caption{$H_{0+}$: $(I_{N\pi},I_J,I_N)=(\frac{3}{2},1,\frac{1}{2})$}
        \end{subfigure} &
        	\begin{subfigure}{0.48\textwidth}
	\centering
	\includegraphics[width=\linewidth,angle=0]{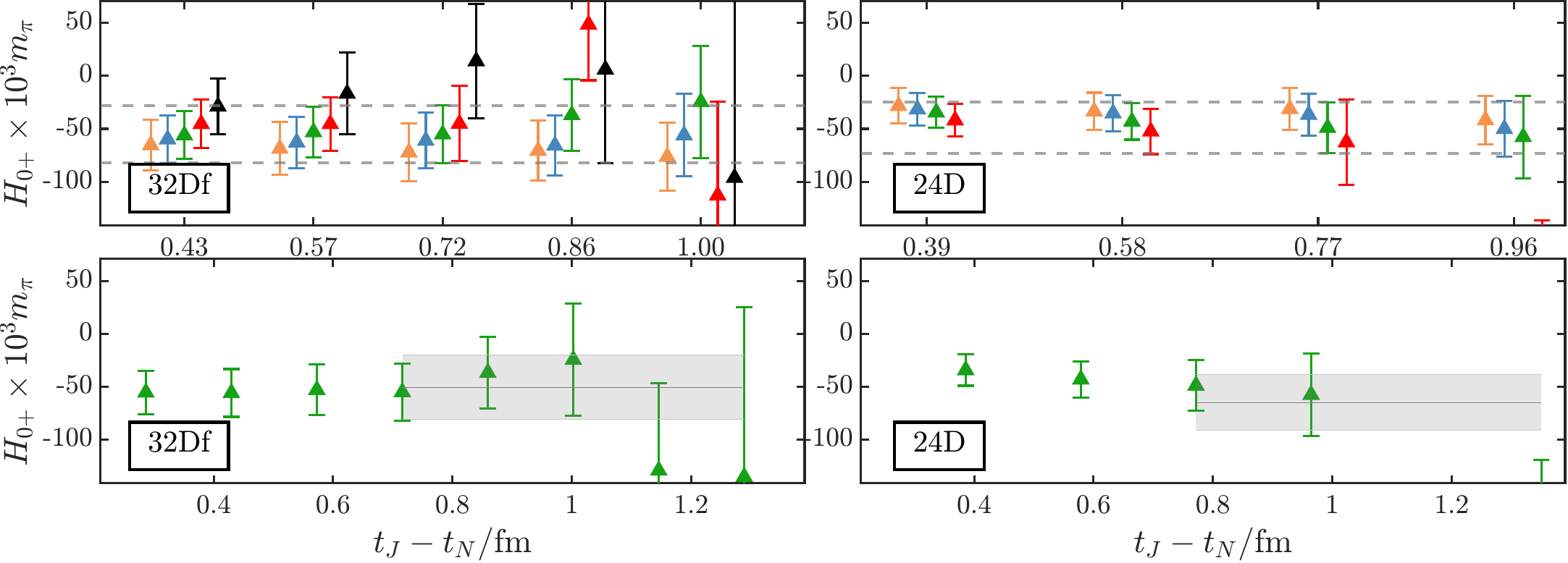}
	\caption{$H_{0+}$: $(I_{N\pi},I_J,I_N)=(\frac{1}{2},1,\frac{1}{2})$}
         \end{subfigure}\\
       \begin{subfigure}{0.48\textwidth}
	\centering
	\includegraphics[width=\linewidth,angle=0]{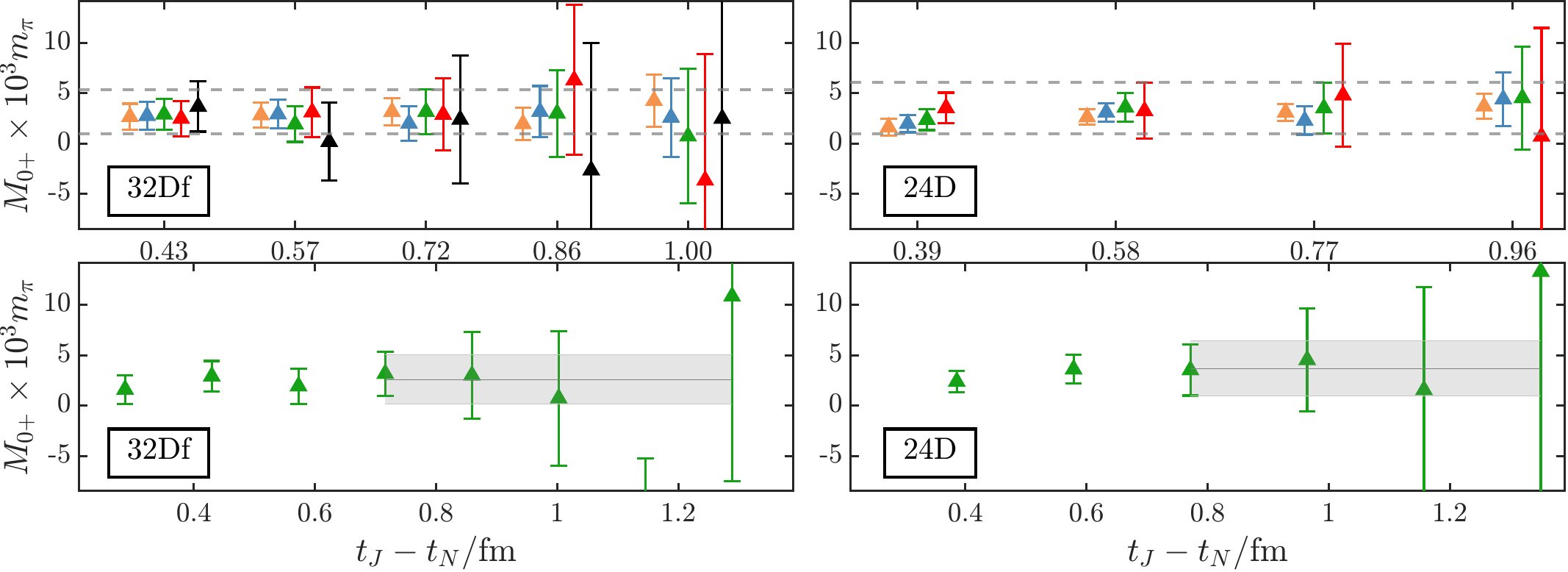}
	\caption{$M_{0+}$: $(I_{N\pi},I_J,I_N)=(\frac{3}{2},1,\frac{1}{2})$}
       \end{subfigure} &
       \begin{subfigure}{0.48\textwidth}
	\centering
	\includegraphics[width=\linewidth,angle=0]{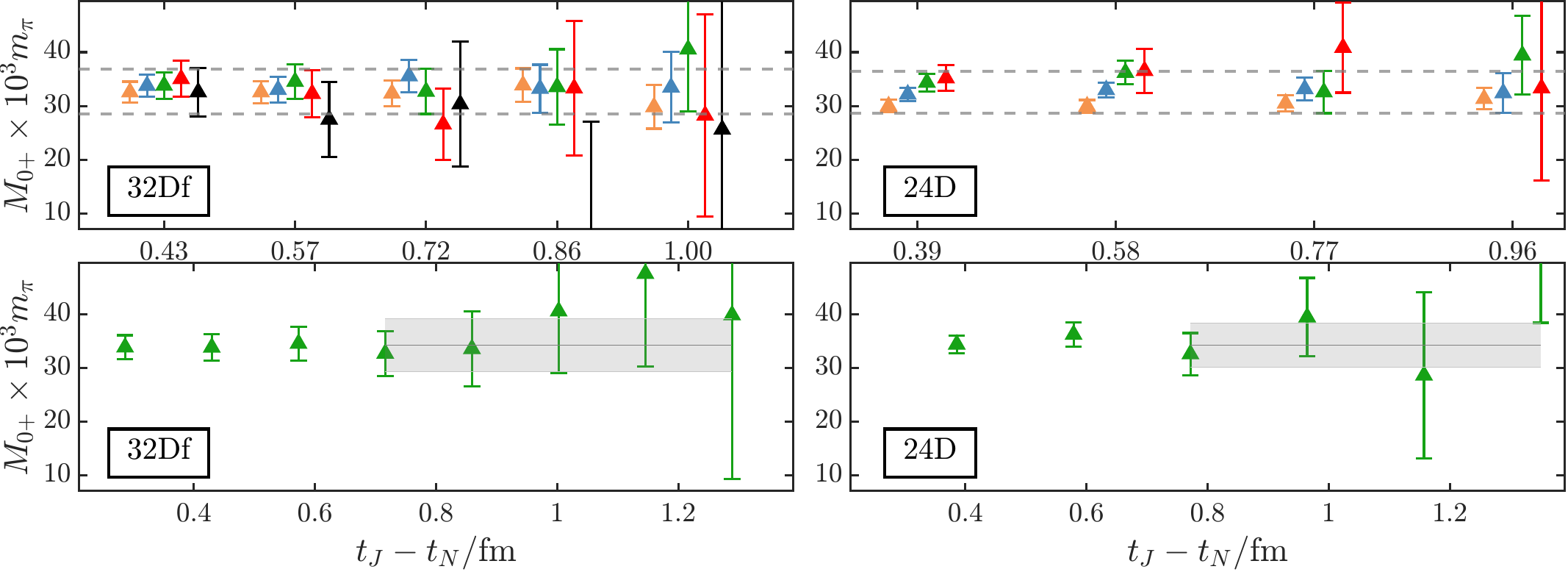}
	\caption{$M_{0+}$: $(I_{N\pi},I_J,I_N)=(\frac{1}{2},1,\frac{1}{2})$}
       \end{subfigure}
	\end{tabular}
	\caption{Fit of various multipole amplitudes for different isospin channels, using the momentum mode $(0,0,1)$ as an example.}
	\label{fig:multipole_amplitude_fit}
	\end{figure*}

	Fig.~\ref{fig:multipole_amplitude_fit} consists of twelve subfigures, all constructed using GEVP-corrected data for both the $N\pi$ and $N$ sides. 
	Each subfigure contains four plots: the two on the left correspond to 32Df, while the two on the right correspond to 24D. In the upper plots, the $x$-axis represents the time separation $t_J-t_N$ in units of fm.
	For each fixed $t_J-t_N$, the size of the fitting window for $t_{N\pi}-t_J$, defined as $t_{\mathrm{max}} - t_{\mathrm{min}}$ (where $t_{\mathrm{min}}$ and $t_{\mathrm{max}}$ 
	are the starting and ending points of the fitting window), is set to 0.57 fm for 32Df and 0.58 fm for 24D. 
	We then vary $t_{\mathrm{min}}$ to examine whether the chosen window effectively controls excited-state contamination. 
	Different values of $t_{\mathrm{min}}$ are represented by different colors. For 32Df, $t_{\mathrm{min}}$ ranges from 0.43 to 1.00 fm, and for 24D, it ranges from 0.39 to 0.96 fm.
	The results at $t_J-t_N=0.72$ fm and $t_{\mathrm{min}}=0.72$ fm for 32Df, as well as those at $t_J-t_N=0.77$ fm and $t_{\mathrm{min}}=0.77$ fm for 24D (indicated by the dashed lines),
	are found to be generally consistent with those obtained using smaller or larger $t_{\mathrm{min}}$ values. Thus, we consider the $t_{\mathrm{min}}$ values 
	marked by the dashed lines to be the optimal choices for the fit. The corresponding fitting results as a function of $t_J-t_N$ are shown in the lower plots, 
	where a second fit is performed to extract the final multipole amplitude results, presented in Figs.~\ref{fig:comparison_phy_basis} and \ref{fig:isospin_basis}.

\end{document}